\documentclass[pra,twocolumn]{revtex4-1}
\usepackage{hyperref,graphicx,subfig,color,filecontents,soul,physics,mathrsfs,relsize,amsmath,colortbl,tabularx,stackengine,dcolumn,bm,float,amssymb,tikz,accents} 

\usepackage{tikz}
\usetikzlibrary{arrows}
\usepackage[utf8]{inputenc}
\usetikzlibrary{arrows.meta}
\definecolor{blue1}{rgb}{0.61, 0.87, 1.0}
\definecolor{blue2}{rgb}{0.6, 0.4, 0.8}
\definecolor{blue3}{rgb}{0.0, 0.44, 1.0}
\definecolor{blue4}{rgb}{0.54, 0.17, 0.89}
\definecolor{blue5}{rgb}{0.01, 0.28, 1.0}
\definecolor{green1}{rgb}{0.0, 0.42, 0.24}
\definecolor{green1}{rgb}{0.01, 0.75, 0.24}
\usetikzlibrary{decorations.pathmorphing}
\definecolor{Redhawks-Red}{RGB}{200,16,46}

\definecolor{LightCyan}{rgb}{0.88,1,1}
\definecolor{corn}{rgb}{0.98, 0.93, 0.36}
\definecolor{pastelyellow}{rgb}{0.99, 0.99, 0.59}
\definecolor{DarkGreen}{rgb}{0,0.6,0.2}

\newcommand{\be}{\begin{equation}}
\newcommand{\ee}{\end{equation}}
\newcommand{\bea}{\begin{eqnarray}}
\newcommand{\eea}{\end{eqnarray}}
\newcommand*{\myeqref}[2][Eq.~]{%
  \hyperref[{#2}]{#1(\ref*{#2})}%
}
\def\equationautorefname#1#2\null{%
  Eq.#1(#2\null)%
}

\begin{document}
\title{Chirality-assisted enhancement of tripartite entanglement in waveguide QED}
\author{Logan Patrick$^\ast$} \author{Umar Arshad$^\ast$} \author{Dingyu Guo}
\thanks{These authors have contributed equally to this work.}
\author{Imran M. Mirza}
\email{mirzaim@miamioh.edu}
\affiliation{Macklin Quantum Information Sciences,\\ Department of Physics, Miami University, Oxford, Ohio 45056, USA}

\begin{abstract}
We study the generation and control of genuine tripartite entanglement among quantum emitters (QEs) that are side coupled to one-dimensional spin-momentum locked (or chiral) waveguides. By applying the machinery of Fock state master equations along with the recently proposed concurrence fill measure of tripartite entanglement [S. Xie and J. H. Eberly, Phys. Rev. Lett. 127, 040403 (2021)], we analyze how three-photon Gaussian wavepackets can distribute entanglement among two and three QEs. We show that with a five times larger waveguide decay rate in the right direction as compared to the left direction, the maximum value of tripartite entanglement can be elevated by $35\%$ as compared to the symmetric scenario where both left and right direction decay rates are equal. Additionally, chirality can maintain the tripartite entanglement for longer times in comparison to the corresponding symmetric decay rate situation. Finally, we study the influence of detunings and spontaneous emission on the resulting entanglement. We envision quantum networking and long-distance quantum communication as two main areas of applications of this work.
\end{abstract}

\maketitle

\section{Introduction}
A relatively recent way to accomplish optical nonreciprocity at the quantum level is to utilize the so-called ``spin-momentum locking of light" \cite{bliokh2015spin, aiello2015transverse}. To understand this phenomenon, one can consider an ultra-thin (subwavelength) dielectric waveguide or tapered nanofiber guiding an s-polarized electromagnetic wave propagating in the $x-$direction. Upon reaching the tapered region of the fiber, some of the light field evanescently leaks out from the fiber with an exponentially decaying amplitude in the direction perpendicular to the propagation (say $y-$direction). Due to the symmetry of the problem, the resultant electric field vector $\textbf{E}(x,y)$ can be expressed in the following fashion 
\begin{equation}
\textbf{E}(x,y)=\textbf{E}_{0x}e^{ikx}e^{-\beta y}+\textbf{E}_{0y}e^{ikx}e^{-\beta y}.
\end{equation}
Here $\textbf{E}_{0x}$ and $\textbf{E}_{0y}$ refer to the electric field components in the $x$ and $y$-direction, respectively, $k$ is the wavenumber, and $\beta$ represents the inverse decay length. The relation between the electric field components can be established through the use of Maxwell's equations yielding 
\begin{equation}
\textbf{E}_{0x}=-\frac{i\beta}{k}\textbf{E}_{0y}.
\end{equation}
For air-glass interfaces with $\beta\approx k$ \cite{poudyal2020single}, the longitudinal and transverse components of the electric field turn out to be directly proportional with a phase difference of $-\pi/2$, i.e. $\textbf{E}_{0x}\approx -i \textbf{E}_{0y}$. Thus the resultant electric field carries an electric field component oscillating in the propagation direction consequently breaking the often encountered transverse nature of light propagation. This remarkable feature can then be quantified with the use of a ``spin vector" whose magnitude represents the deviation from a perfect circular polarization -- a magnitude of one (zero) referring to circular (linear) polarization, while the direction shows the net polarization direction. One of the key consequences of this effect is the flipping of polarization direction of the spin vector (from out of the page to into the page) with the reversal of light propagation direction in the nanofiber (from left-to-right to right-to-left), hence leading to the phenomenon of spin-momentum locking of light \cite{junge2013strong, scheucher2021cavity}. 

\begin{figure*}
\centering
\includegraphics[width=0.75\textwidth]{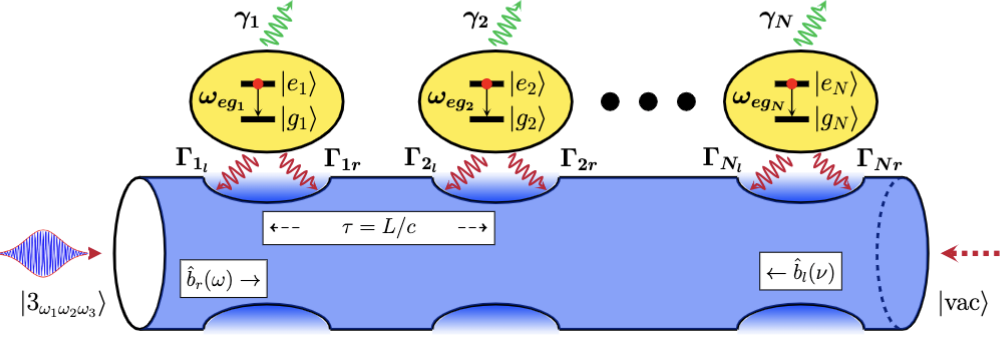}
\captionsetup{
  format=plain,
  margin=1em,
  justification=raggedright,
  singlelinecheck=false
}
\caption{(Color online) The system studied in this work: A chain of two-level quantum emitters side coupled to a one-dimensional waveguide which is driven by a three-photon wavepacket (represented by the state $\ket{3_{\omega_1\omega_2\omega_3}}$) from the left end of the waveguide. The right end of the waveguide is not driven by any photon source and is, therefore, labeled with a vacuum state $\ket{vac}$. The valleys on the waveguide surface are drawn to indicate the tapered region of the nanofiber where the QEs are trapped to accomplish chiral light-matter interactions. $\gamma_j$ represents the non-waveguide (or spontaneous) emission rate for the $j$th QE. For further details about the system parameters, see the text below. }\label{Fig1}
\end{figure*}

One can then imagine the fascinating consequences offered by the spin-momentum locking of light in a variety of light-matter interfaces. To emphasize one such possibility, we consider a two-level QE coupled with such a one-dimensional chiral waveguide. The rate of emission from such a QE in the right ($r$) and left ($l$) directions of the waveguide is known to follow
\begin{equation}
\Gamma_{r/l} \propto | \textbf{d}^{*}\cdot \textbf{E}_{r/l}|^2,
\end{equation}
where $\textbf{d}^{*}$ is the conjugate of the transition dipole matrix element and $\textbf{E}_{r/l}$ is the electric field vector of the right and left modes. Utilization of a spin-momentum locked waveguide alongwith special type of QEs with polarization-dependent transitions then one can see that only a proper match of the polarizations in both QE and the waveguide will result in the emission/absorption of photon. This polarization-dependent preferential photon absorption/emission has recently opened a new research area called ``chiral quantum optics" \cite{lodahl2017chiral}.

In the last five years or so chiral quantum optics, has been achieved in various physical platforms \cite{balykin2004atom, solano2017super, scarpelli201999} and the area has witnessed a variety of novel effects \cite{mirza2017chirality, mirza2018influence, mirza2020dimer, poudyal2020collective, yan2018targeted, li2018quantum}. For instance, we and others have studied emitter-emitter entanglement dynamics in chiral and non-chiral wQED \cite{gonzalez2015chiral, mok2020long, buonaiuto2019dynamical}. In particular, we have shown that for strongly coupled single and two-photon wQED cases such chiral light-matter interaction can lead to enhancement in the maximum value of bipartite emitter-emitter entanglement by a factor of 3/2 and 2, respectively as compared to the corresponding non-chiral coupling case \cite{mirza2016multiqubit, mirza2016two}. 

In this work, we examine the novel problem of genuine tripartite entanglement generation and control among up to three QEs simultaneously coupled with bidirectional and chiral waveguides. Worthwhile to emphasize here is the fact that the discussion of the tripartite entanglement should not be treated as a straightforward extension of single or two-photon entanglement problems, but by doing so we enter the richer and more challenging domain of multipartite entanglement \cite{szalay2015multipartite, m2019tripartite} where tripartite entanglement can serve as the simplest case study. As the theoretical tools, we work within the framework of Fock state master equations \cite{gheri1998photon, baragiola2012n, patrick2023fock} and calculate the genuine tripartite entanglement among up to three two-level QEs using concurrence and concurrence fill criteria \cite{xie2021triangle}. As some of the main findings, we find that, as compared to the corresponding non-chiral (symmetric bidirectional) models, the chirality (five times larger emission rate into the right direction in the waveguide as compared to the left direction) can raise the maximum tripartite entanglement value by 35\% of by a factor of $\sim 5/14$. Additionally, for both on-resonant and off-resonant cases, chirality aids to maintain tripartite for a longer duration as compared to the symmetric bidirectional problem. Furthermore, chirality also exhibits better robustness against spontaneous emission compared to non-chiral scenarios. 

The rest of the paper is structured as follows. In Sec.~\ref{sec:II} we discuss the theoretical description of our system. In Sec.~\ref{sec:III}, we introduce the entanglement measure and discuss our results. In Sec.~\ref{sec:IV} we close with a summary and point out possible future directions. Finally, in the Appendices we outline the derivation of the three-photon Fock-state master equation for cascaded multi-emitter wQED.  

\section{Theoretical Description}\label{sec:II}
\subsection{Model}
\vspace{-2mm}
As shown in Fig.~\ref{Fig1}, our system consists of a chain of two-level QEs (qubits, quantum dots, artificial atoms, natural atoms, etc.) side coupled to a bidirectional dispersionless and lossless waveguide (tapered fiber). The free Hamiltonian of the emitter chain is given by
\begin{align}\label{Hsys}
\hat{\mathcal{H}}_{QE}=\hbar\sum\limits^{N}_{j=1}\widetilde{\Delta}_j\hat{\sigma}^\dagger_j\hat{\sigma}_j,  
\end{align}
where $\widetilde{\Delta}_j=\omega_{eg_{j}}-\omega_p-i\gamma_j$ is the detuning between the transition frequency $\omega_{eg_{j}}$ of the $j$th QE and the peak frequency $\omega_{p}$ of the three-photon wavepacket. Note that in our model there is no direct coupling (such as dipole-dipole interaction) present among the QEs rather the interaction is mediated through the waveguide field. $\hat{\sigma}_j\equiv\ket{g_j}\bra{e_j}$ is the standard lowering operator for the $j$th QE with $\ket{g_j}(\ket{e_j})$ being the ground (excited) state. The QE raising and lowering operators follow the standard Ferminonic commutation relation: $\left\lbrace \hat{\sigma}_i,\hat{\sigma}^\dagger_j \right\rbrace=\delta_{ij}$. 

Next, we model the waveguide as a collection of two independent multimode quantum harmonic oscillators, one for the left ($l$) direction, and the other for the right ($r$) direction. The corresponding photon annihilation operators are labelled as $\hat{b}_l(\nu)$ and $\hat{b}_r(\omega)$ for the $\nu$th and $\omega$th mode. These operators follow the typical Bosonic commutation relations: $[\hat{b}_r(\omega),\hat{b}^\dagger_r(\omega^{'})]=\delta(\omega-\omega^{'})$ and $[\hat{b}_l(\nu),\hat{b}^\dagger_l(\nu^{'})]=\delta(\nu-\nu^{'})$. Thus, the waveguide Hamiltonian $\hat{\mathcal{H}}_{w}$ takes the form
\begin{align}\label{Hbath}
\hat{\mathcal{H}}_{w}=\hbar\int\limits^{+\infty}_{-\infty}\omega\hat{b}^\dagger_r(\omega)\hat{b}_r(\omega)d\omega + \hbar\int\limits^{+\infty}_{-\infty}\nu\hat{b}^\dagger_l(\nu)\hat{b}_l(\nu)d\nu.
\end{align}
In $\hat{\mathcal{H}}_{w}$, we have considered an infinitely large number of closely spaced waveguide modes such that the integration over all modes is justified. Finally, under the rotating wave approximation, the interaction between the QEs and waveguide field is described in the following Hamiltonian 
\begin{align}\label{Hint}
\hat{\mathcal{H}}_{int} =& -i\hbar\sum\limits^N_{j=1}\Big[\int\limits^{+\infty}_{-\infty} \sqrt{\frac{\Gamma_{jr}}{2\pi}}e^{ik_0d_j}\hat{\sigma}^\dagger_j\hat{b}_r(\omega)d\omega\nonumber\\
&+ \int\limits^{+\infty}_{-\infty} \sqrt{\frac{\Gamma_{jl}}{2\pi}}e^{-ik_0d_j}\hat{\sigma}^\dagger_j\hat{b}_l(\nu) d\nu \Big]+h.c.,
\end{align}
where we have assumed $\Gamma_{jr}(\omega)\approx\Gamma_{jr}(\omega_{eg_{j}})\equiv\Gamma_{jr}$ and $\Gamma_{jl}(\omega)\approx\Gamma_{jl}(\omega_{eg_{j}})\equiv\Gamma_{jl}$. Note that in this assumption we have not applied the Markov approximation (flat bath spectrum around the system resonance) \cite{gardiner2004quantum} rather we are considering a highly localized interaction. $d_j$ represents the location of the $j$th emitter with $d_{j+1}-d_{j}=L$ being the separation between two consecutive QEs (or lattice constant) that correspond to the time delay $\tau=L/c$. The parameter $k_0=\omega_{eg}/c$ is the wavenumber associated with the atomic transition frequency, while $c$ represents the group velocity in the waveguide. The net Hamiltonian of the global system (QEs, waveguide, and their interaction) is given by $\hat{\mathcal{H}}=\hat{\mathcal{H}}_{QE} + \hat{\mathcal{H}}_{w} + \hat{\mathcal{H}}_{int}$. 

\subsection{Driven dissipative dynamics}
As shown in Fig.~\ref{Fig1} the left end of our wQED setup is driven by a reservoir that initially exists in a three-photon Fock state which is unlike the standard studied scenario where a classical coherent light source drives the system. Keeping in view this important distinction, in Appendices we derive the master equation apt for the present problem and study the driven dissipative dynamics of our wQED setup through such a bi-directional three-photon Fock state master equation which is reported as follows 
\vspace{-12mm}
\allowdisplaybreaks
\begin{widetext}
\begin{subequations}\label{3PFME}
\begin{align}
&\frac{d\hat{\rho}_{3,3}(t)}{dt}=\hat{\mathcal{L}}\left[\hat{\rho}_{3,3}\right]+\sum^N_{i=1}\sqrt{\Gamma_{ir}}\left(\sqrt{3}e^{ik_0d_i}g(t)\left[\hat{\rho}_{2,3},\hat{\sigma}^\dagger_i\right]+\sqrt{3}e^{-ik_0d_i}g^\ast(t)\left[\hat{\sigma}_i,\hat{\rho}^\dagger_{2,3}\right]\right),\\
&\frac{d\hat{\rho}_{2,3}(t)}{dt}=\hat{\mathcal{L}}\left[\hat{\rho}_{2,3}\right]+\sum^N_{i=1}\sqrt{\Gamma_{ir}}\left(\sqrt{2}e^{ik_0d_i}g(t)\left[\hat{\rho}_{1,3},\hat{\sigma}^\dagger_i\right]+\sqrt{3}e^{-ik_0d_i}g^\ast(t)\left[\hat{\sigma}_i,\hat{\rho}_{2,2}\right]\right),\\
&\frac{d\hat{\rho}_{1,3}(t)}{dt}=\hat{\mathcal{L}}\left[\hat{\rho}_{1,3}\right]+\sum^N_{i=1}\sqrt{\Gamma_{ir}}\left(e^{ik_0d_i}g(t)\left[\hat{\rho}_{0,3},\hat{\sigma}^\dagger_i\right]+\sqrt{3}e^{-ik_0d_i}g^\ast(t)\left[\hat{\sigma}_i,\hat{\rho}^\dagger_{1,2}\right]\right),\\
&\frac{d\hat{\rho}_{0,3}(t)}{dt}=\hat{\mathcal{L}}\left[\hat{\rho}_{0,3}\right]+\sum^N_{i=1}\sqrt{3\Gamma_{ir}}e^{-ik_0d_i}g^\ast(t)\left[\hat{\sigma}_i,\hat{\rho}^\dagger_{0,2}\right],\\
&\frac{d\hat{\rho}_{2,2}(t)}{dt}=\hat{\mathcal{L}}\left[\hat{\rho}_{2,2}\right]+\sum^N_{i=1}\sqrt{\Gamma_{ir}}\left(\sqrt{2}e^{ik_0d_i}g(t)\left[\hat{\rho}_{1,2},\hat{\sigma}^\dagger_i\right]+\sqrt{2}e^{-ik_0d_i}g^\ast(t)\left[\hat{\sigma}_i,\hat{\rho}^\dagger_{1,2}\right]\right),\\
&\frac{d\hat{\rho}_{1,2}(t)}{dt}=\hat{\mathcal{L}}\left[\hat{\rho}_{1,2}\right]+\sum^N_{i=1}\sqrt{\Gamma_{ir}}\left(e^{ik_0d_i}g(t)\left[\hat{\rho}_{0,2},\hat{\sigma}^\dagger_i\right]+\sqrt{2}e^{-ik_0d_i}g^\ast(t)\left[\hat{\sigma}_i,\hat{\rho}_{1,1}\right]\right),\\
&\frac{d\hat{\rho}_{0,2}(t)}{dt}=\hat{\mathcal{L}}\left[\hat{\rho}_{0,2}\right]+\sum^N_{i=1}\sqrt{2\Gamma_{ir}}e^{-ik_0d_i}g^\ast(t)\left[\hat{\sigma}_i,\hat{\rho}_{0,1}\right],\\
&\frac{d\hat{\rho}_{1,1}(t)}{dt}=\hat{\mathcal{L}}\left[\hat{\rho}_{1,1}\right]+\sum^N_{i=1}\sqrt{\Gamma_{ir}}\left(e^{ik_0d_i}g(t)\left[\hat{\rho}_{0,1},\hat{\sigma}^\dagger_i\right]+e^{-ik_0d_i}g^\ast(t)\left[\hat{\sigma}_i,\hat{\rho}^\dagger_{0,1}\right]\right),\\
&\frac{d\hat{\rho}_{0,1}(t)}{dt}=\hat{\mathcal{L}}\left[\hat{\rho}_{0,1}\right]+\sum^N_{i=1}\sqrt{\Gamma_{ir}}e^{-ik_0d_i}g^\ast(t)\left[\hat{\sigma}_i,\hat{\rho}_{0,0}\right],\\
&\frac{d\hat{\rho}_{0,0}(t)}{dt}=\hat{\mathcal{L}}\left[\hat{\rho}_{0,0}\right].
\end{align}
\end{subequations}
\end{widetext}
Here we would also like to point out that a similar Fock-state master equation valid for $N$ photons has also been reported in the past (see for example Eq.~(21) in \cite{baragiola2012n}). However, there are two main differences between our three-photon Fock-state master equation and the one reported in Ref.~\cite{baragiola2012n}. One is the absence of the terms in our master equation that are quadratic in $g(t)$ which are known to appear in the case of nonlinear interactions (for instance, in cavity quantum optomechanics \cite{aspelmeyer2014cavity}) or in the case of adiabatically eliminated multi-level quantum systems \cite{brion2007adiabatic}. Since our problem doesn't address both of these scenarios, therefore, the absence of such quadratic terms in Eq.~(\ref{3PFME}) is understandable. The second difference stems from the fact that, unlike the master equation reported in Ref.~\cite{baragiola2012n}), our master equation incorporates bidirectional couplings between QEs and photon wavepacket which is suitable to study wQED problems). 

The Liouvillian superoperator $\hat{\mathcal{L}}$ appearing in the aforementioned equation set (\ref{3PFME}) and applied to an operator $\hat{\varrho}$ consists of three parts
\begin{align}
\hat{\mathcal{L}}[\hat{\varrho}]=\hat{\mathcal{L}}_{cs}[\hat{\varrho}]+\hat{\mathcal{L}}_{pd}[\hat{\varrho}]+\hat{\mathcal{L}}_{cd}[\hat{\varrho}],
\end{align}
with $\hat{\mathcal{L}}_{cs}[\hat{\varrho}]$, $\hat{\mathcal{L}}_{pd}[\hat{\varrho}]$, and $\hat{\mathcal{L}}_{cd}[\hat{\varrho}]$ respectively represent the closed system dynamics, pure decay of energy from the system into the environmental degrees of freedom, and cooperative decay due to collective QE effects. These Liouvillian subparts are given by 
\begin{subequations}
\begin{align}
&\hat{\mathcal{L}}_{cs}[\hat{\varrho}]\equiv \frac{-i}{\hbar}\left[\hat{\mathcal{H}}_{QE}, \hat{\varrho}\right],\\
&\hat{\mathcal{L}}_{pd}[\hat{\varrho}]\equiv -\sum\limits^N_{i=1}\Gamma_{irl}\left(\hat{\sigma}^\dagger_i\hat{\sigma}_i\hat{\varrho}-2\hat{\sigma}_i\hat{\varrho}\hat{\sigma}^\dagger_i+\hat{\varrho}\hat{\sigma}^\dagger_i\hat{\sigma}_i\right),\\
& \hat{\mathcal{L}}_{cd}[\hat{\varrho}]\equiv-\sum\limits^N_{i\neq j=1}\left(\sqrt{\Gamma_{ir}\Gamma_{jr}}~\delta_{i>j}+\sqrt{\Gamma_{il}\Gamma_{jl}}~\delta_{i<j}\right)\nonumber\\
&\hspace{14mm}\times\left\lbrace(\hat{\sigma}^\dagger_i\hat{\sigma}_j\hat{\varrho}-\hat{\sigma}\hat{\varrho}\hat{\sigma}^\dagger_j)e^{-2\pi iD(i-j)}-h.c.\right\rbrace,
\end{align}
\end{subequations}
where $2\Gamma_{irl}=\Gamma_{ir}+\Gamma_{il}$. The Kronecker delta functions appearing in the expression of $\hat{\mathcal{L}}_{cd}[\hat{\varrho}]$ are defined as $\delta_{i\gtrless j}=1$, $\forall$ $i\gtrless j$. The parameter $D$ represent the ratio of inter-emitter separation $L$ and the resonant wavelength $\lambda_0$ i.e. $\lambda_0=2\pi c/\omega_{eg}$. Finally, the explicit form of the various operators appearing in Eq.~\eqref{3PFME} are given by
\begin{subequations}
\begin{align}
&\hat{\rho}_{3,3}(t)=\tr_R\left\lbrace \hat{U}(t;t_0)\hat{\rho}_s(t_0)\ket{\Psi_r}\bra{\Psi_r}\hat{\rho}_{l}(t_0)\hat{U}^\dagger(t;t_0)\right\rbrace,\\
&\hat{\rho}_{j,3}(t)=\tr_R\left\lbrace \hat{U}(t;t_0)\hat{\rho}_s(t_0)\ket{\Psi_j}\bra{\Psi_r}\hat{\rho}_{l}(t_0)\hat{U}^\dagger(t;t_0)\right\rbrace,\\
&\hat{\rho}_{j,2}(t)=\tr_R\left\lbrace \hat{U}(t;t_0)\hat{\rho}_s(t_0)\ket{\Psi_j}\bra{\Psi_2}\hat{\rho}_{l}(t_0)\hat{U}^\dagger(t;t_0)\right\rbrace,\\
&\hat{\rho}_{k,1}(t)=\tr_R\left\lbrace \hat{U}(t;t_0)\hat{\rho}_s(t_0)\ket{\Psi_k}\bra{\Psi_1}\hat{\rho}_{l}(t_0)\hat{U}^\dagger(t;t_0)\right\rbrace,\\
&\hat{\rho}_{0,0}(t)=\tr_R\left\lbrace \hat{U}(t;t_0)\hat{\rho}_s(t_0)\ket{vac}\bra{vac}\hat{\rho}_{l}(t_0)\hat{U}^\dagger(t;t_0)\right\rbrace,
\end{align}
\end{subequations}
here $ j=2,1,0,$, $k=1,0$, $\hat{U}(t;t_0)$ is the time evolution operator, and $\hat{\rho}_l(t)$ is the density operator for the left continuum in the waveguide. $\ket{\Psi_r}$, $\ket{\Psi_2}$ and $\ket{\Psi_1}$ are the three-, two- and one-photon reservoir states, respectively with $\ket{\Psi_0}=\ket{vac}$. Note that in the above-mentioned set of operators, only the diagonal operators can be categorized as physically valid density matrices. The rest of the off-diagonal operators are not density matrices but they do obey a useful property that $\hat{\rho}^\dagger_{j,3}(t) = \hat{\rho}_{3,j}(t)$, $\hat{\rho}^\dagger_{j,2}(t) = \hat{\rho}_{2,j}(t)$, and $\hat{\rho}^\dagger_{k,1}(t)=\hat{\rho}_{1,k}(t)$.

\subsection{Initial conditions}
Initially, we consider all QEs to be in their ground state with the right waveguide continuum in a three-photon wavepacket with the joint spectral density function $\mathcal{G}(\omega_1,\omega_2,\omega_3)$ and the left continuum in a vacuum state i.e. the initial pure state $\ket{\Psi}$ of the system and environment takes the form
\begin{align}
&\ket{\Psi}=\ket{\Psi_{QE}}\otimes\ket{\Psi_r}\otimes\ket{\Psi_l}= 
\bigotimes_j\ket{g_j}\otimes\ket{\Psi_r}\otimes\ket{vac},\nonumber\\
&\textit{with}~ \ket{\Psi_{r}}=\frac{1}{\sqrt{3!}}\int\limits^{+\infty}_{-\infty}\int\limits^{+\infty}_{-\infty}\int\limits^{+\infty}_{-\infty} d\omega_1 d\omega_2 d\omega_3 ~\mathcal{G}(\omega_1,\omega_2,\omega_3)\nonumber\\
&\hspace{35mm}\times\hat{b}^\dagger_r(\omega_1)\hat{b}^\dagger_r(\omega_2)\hat{b}^\dagger_r(\omega_2)\ket{vac}.
\end{align}
At this stage we keep the form of $\mathcal{G}(\omega_1,\omega_2,\omega_3)$ general, however, the normalization condition on $\ket{\Psi}$ requires any $\mathcal{G}(\omega_1,\omega_2,\omega_3)$ must follow the condition
\begin{align}
\int\limits^{+\infty}_{-\infty}\int\limits^{+\infty}_{-\infty}\int\limits^{+\infty}_{-\infty} \left|\mathcal{G}(\omega_1,\omega_2,\omega_3)\right|^2 d\omega_1 d\omega_2 d\omega_3=1,
\end{align}
where in arriving at this condition we have assumed the spectral function $\mathcal{G}(\omega_1,\omega_2,\omega_3)$ is symmetric under the exchange of mode frequencies $\omega_1$, $\omega_2$, and $\omega_3$. Finally, we impose
\begin{align}\label{inirhos}
&\hat{\rho}_{m,m}(0)=\hat{\rho}_{sys}(0)=\bigotimes_j\ket{g_j}\bra{g_j}, \forall m~\text{and}~n=0,1,2,3;\nonumber\\
&\text{and}~\hat{\rho}_{m,n}(0)=0,~\text{with}~m\neq n,
\end{align}
which are the initial conditions followed by the operators appearing in Eq.~\ref{3PFME}.


\section{Results and Discussion}\label{sec:III}
\vspace{-3mm}
In this section, by numerically solving our three-photon bidirectional Fock state master equation we address two questions: 
\vspace{-2mm}
\begin{itemize}
    \item How does the incoming three-photon wavepacket excites the QEs, and as a result how does the population evolve in time?
    \vspace{-2mm}
    \item How does the photon absorption \& emission generate entanglement among QEs and how chirality can impact the entanglement manipulation?
\end{itemize}
\vspace{-2mm}
Albeit Eq.~\eqref{3PFME} is valid for any number of QEs, in the following we focus on situations up to 3 QEs. To set the stage we begin with the simplest possible situation of a single QE. 

\vspace{-5mm}
\subsection{\label{sec:level3.A}One QE case and population dynamics}
\begin{figure}
\centering
\includegraphics[width=2.8in, height=2in]{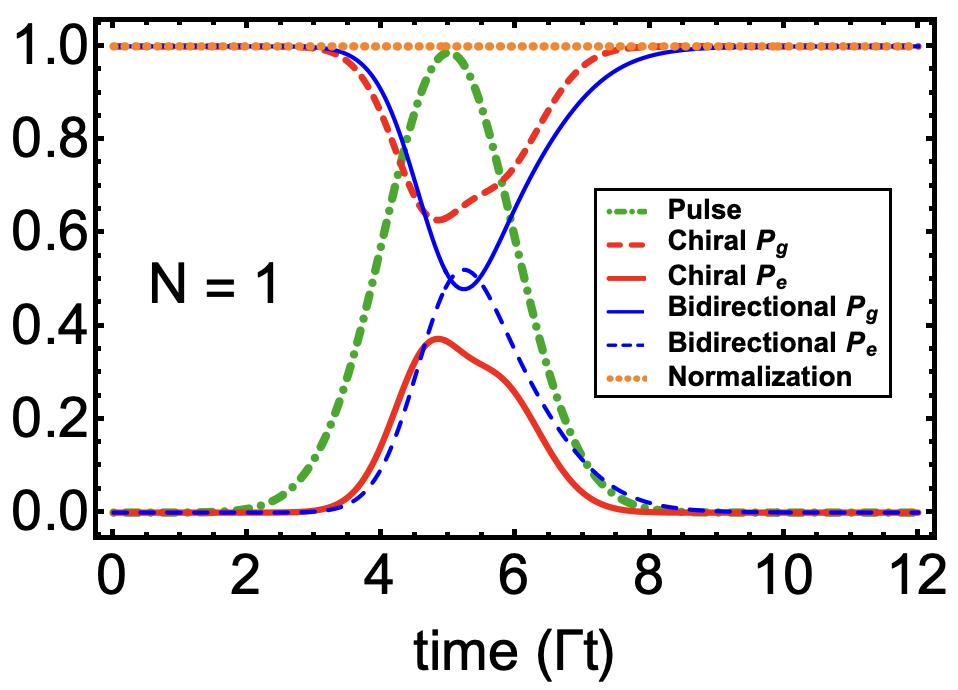} 
\captionsetup{
format=plain,
margin=1em,
justification=raggedright,
singlelinecheck=false
}
\caption{(Color online) Population dynamics, quantified in the units of $\Gamma^{-1}$, for a single ($N=1$) two-level QE when interacted with a three-photon Gaussian wavepacket. We have considered the following common parameters in all curves: $\Delta=0$, $\mu=1.46\Gamma$, $\overline{t}=5\Gamma^{-1}$. For the chiral case, we have set $\Gamma_{1r}/\Gamma_{1l}=5$, while for the bidirectional case, we have selected a symmetric case i.e. $\Gamma_{1r}=\Gamma_{1l}\equiv\Gamma$. The orange dotted horizontal line confirms normalization in both bidirectional and chiral cases.}\label{Fig2}
\end{figure}
\vspace{-3mm}
For the single-QE case (${\rm N=1}$) our free QE Hamiltonian reduces to $\mathcal{\hat{H}}_{QE}=\hbar\widetilde{\Delta}\hat{\sigma}^\dagger\hat{\sigma}$ and as initial conditions we assume $\hat{\rho}_{m,m}(t_0)=\ket{g}\bra{g}$, $\forall m=3,2,1,0$ and the remaining operators to be zero. For the three-photon spectral density function $\mathcal{G}(\omega_1,\omega_2,\omega_3)$ we assume a factorized form such that using Schmidt decomposition we write
\begin{align}
   \mathcal{G}(\omega_1,\omega_2,\omega_3)=\frac{1}{\sqrt{3!}}\sum_{\rm cyc}g_1(\omega_1)g_2(\omega_2)g_3(\omega_2),
\end{align}
where $\sum_{\rm cyc}$ represents the sum over all pairwise cyclic permutation of the indices which counts to a total of 6 terms. We point out that the aforementioned type of decomposition of the spectral density function is experimentally achievable when the three-photon wavepacket is generated by combining the single photons emitted by three independent sources \cite{gheri1998photon, kumar2014controlling}. Moving forward, in all plots to follow we select a real-valued Gaussian temporal profile for each $g$ function i.e. 
\begin{align}
    g(t)=\frac{\sqrt{\mu}}{(2\pi)^{1/4}}\exp\left(-\frac{\mu^2}{4}\left(t-\overline{t}~\right)^2\right).
\end{align}
Here $\mu$ and $\overline{t}$ represent the standard deviation and mean of the Gaussian function, respectively. 

In Fig.~\ref{Fig2} we plot the population dynamics under strong drive condition i.e. $|\Omega^{(max)}(t)|>\Gamma$ with $\Omega(t)=\sqrt{2\Gamma}g(t)$. The rest of the parameters mentioned in the plot caption are selected to generate higher excitation probabilities \cite{wang2011efficient, baragiola2012n}. The green dotted dashed curve shows our three-photon normalized Gaussian wavepacket peaked at $t=\overline{t}=5\Gamma^{-1}$. We have plotted the ground ($P_g$) and excited population ($P_e$) for two cases, namely, a non-chiral or symmetric bidirectional coupling ($\Gamma_{1r}=\Gamma_{1l}$) case (thin blue solid and dashed curves); and a chiral case (thick red solid and dashed curves) in which emission in the right direction is five-time larger than the left direction ($\Gamma_{1r}=5\Gamma_{1l}$). In both cases, we note that as the Gaussian wavepacket begins to interact with the QE, it took almost $t\sim\Gamma^{-1}$ time before the populations begin to change. 

\begin{figure*}
\centering
\begin{tabular}{@{}cccc@{}}
\includegraphics[width=2.35in, height=1.8in]{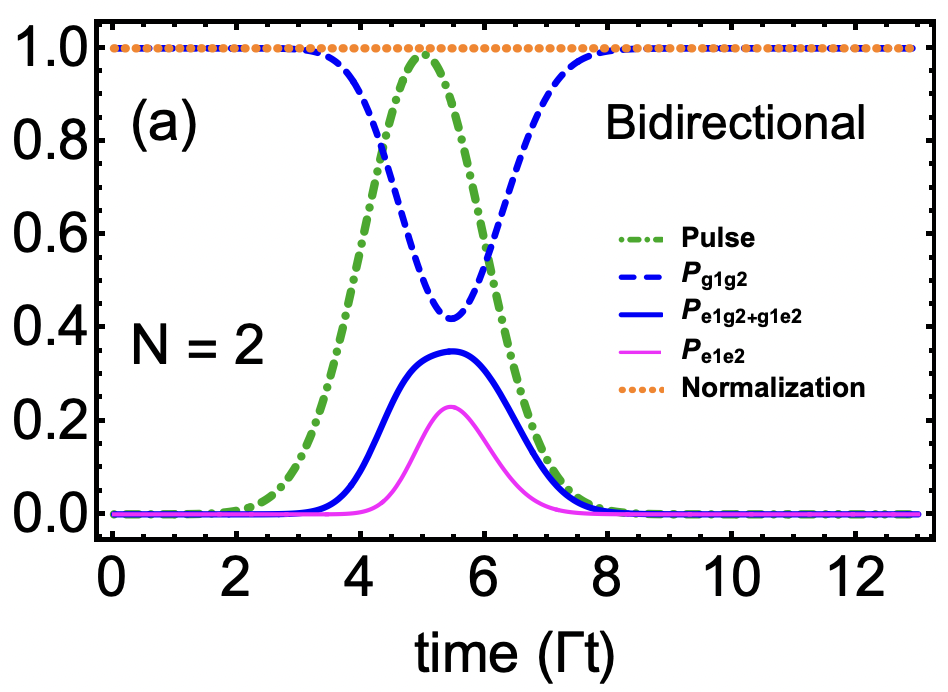} &
\includegraphics[width=2.35in, height=1.8in]{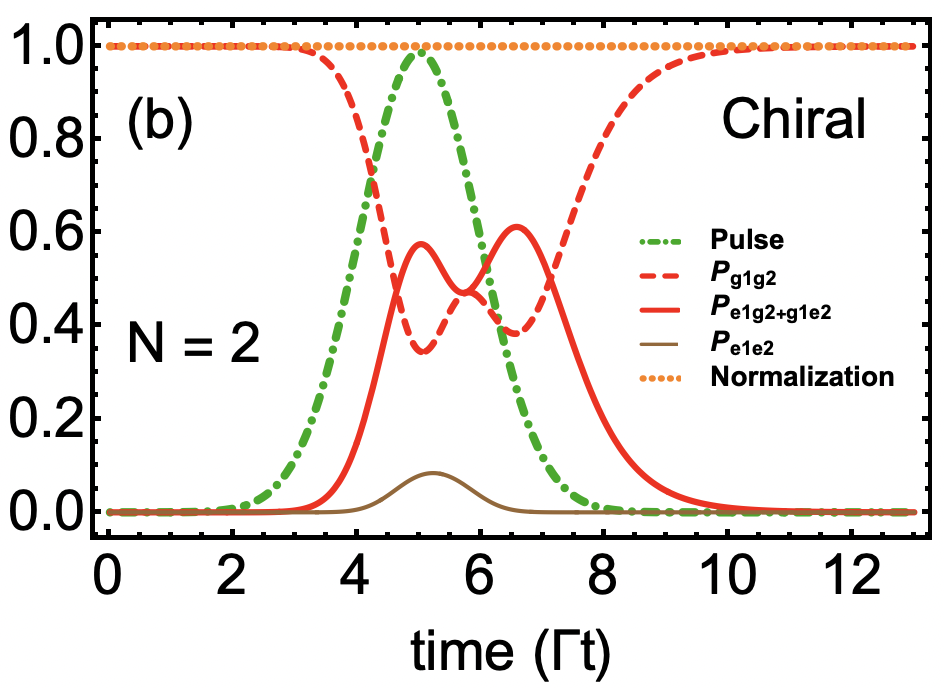}  &
\includegraphics[width=2.35in, height=1.8in]{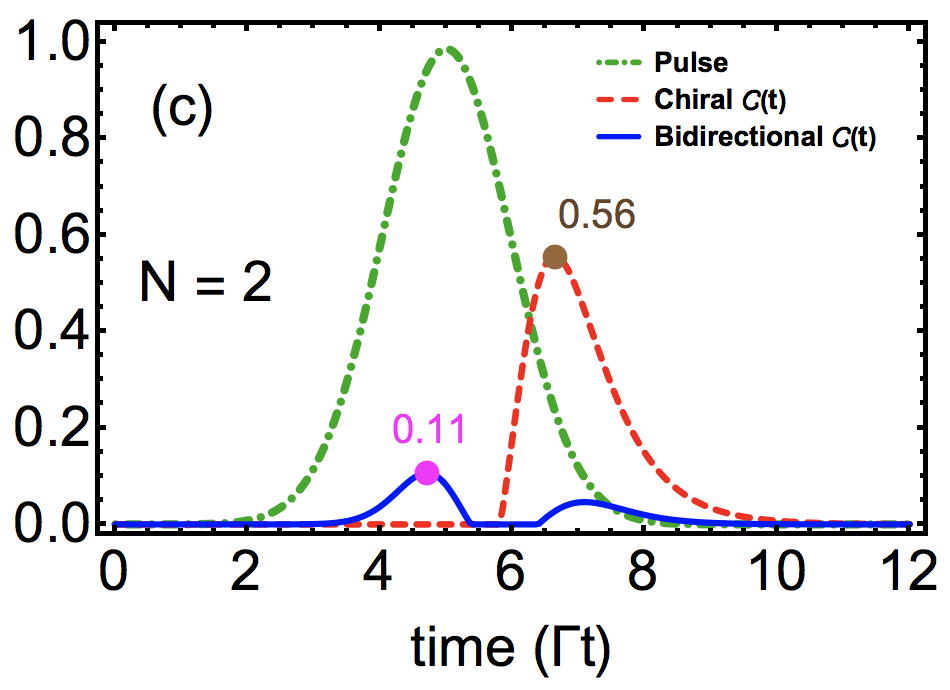}  
\end{tabular}
\captionsetup{
format=plain,
margin=1em,
justification=raggedright,
singlelinecheck=false
}
\caption{(Color online) Two-emitter ${\rm N=2}$ wQED driven by a three-photon Gaussian wavepacket. Population dynamics in (a) Bidirectional case $\Gamma_{ir}=\Gamma_{il}=1$ and (b) Chiral case $\Gamma_{ir}/\Gamma_{il}=5$, $\forall i=1,2$. In the subscripts of $P$ the first and second slots identify the first and second QE, respectively. (c) Entanglement/concurrence evolution in both bidirectional and chiral cases. The location and the maximum value of entanglement have been identified by pink and brown-colored dots. For the sake of simplicity, all QEs are assumed to be identical and time delays have been ignored. The rest of the parameters are the same as used in Fig.~\ref{Fig2}.}\label{Fig3}
\end{figure*}

The maximum value of the excited state probability $P^{(max)}_e$ attained for the bidirectional case turns out to be $0.521$ at $t=5.25\Gamma^{-1}$ which is smaller than the reported value of $0.801$ \cite{baragiola2012n} for the single photon problem due to the involvement of bidirectional decays in our model. Additionally, the shape of $P_e$ follows the profile of Gaussian input which decays as the photon wavepacket leaves the QE region. The chiral case, on the other hand, allowed to attain a smaller value of $P^{(max)}_e=0.373$ due to a higher decay rate into the right waveguide direction. Additionally, this maximum value is achieved at a time $t=4.85\Gamma^{-1}$ slightly before the $P_e$ reaches its maximum value for the bidirectional case. More importantly, we observe the formation of a side shoulder around $t\sim 5.5\Gamma^{-1}$. Such behavior of $P_e$ in the chiral case is known for the single and two-photon wQED problems \cite{mirza2016multiqubit, mirza2016two} and (as discussed below) will help in better emitter-emitter entanglement generation and control. 

\vspace{-3mm}
\subsection{\label{sec:level3.B}Two-QE case and bipartite entanglement}
We now extend our wQED study to two QEs. In addition to new ways of population distribution, the case of two QEs opens the possibility of generating entanglement between the QEs which we quantify through the well-known concurrence measure \cite{wootters1998entanglement, wootters2001entanglement}. For two particles, say particle $A$ and particle $B$, existing in a bipartite pure or mixed state $\hat{\varrho}_{AB}$, Wootter's concurrence $\mathcal{C}_{A(B)}$ is defined as
\begin{align}\label{concurdef}
\mathcal{C}_{A(B)} = \max\left(0,\sqrt{\lambda_1}-\sqrt{\lambda_2}-\sqrt{\lambda_3}-\sqrt{\lambda_4}\right),  
\end{align}
where eigenvalues of operator $\widetilde{\varrho}_{AB}$, $\lambda_i$, $\forall i=1,2,3,4$ are written in a descending order. $\widetilde{\varrho}_{AB}$ is called the spin-flipped density operator which is related to the system density operator and the Pauli spin-flip operator $\hat{\sigma}_y$ through
\begin{align}
\widetilde{\varrho}_{AB}=\hat{\rho}_{AB}\left(\hat{\sigma}_y\otimes\hat{\sigma}_y\right)\hat{\rho}^\ast_{AB}\left(\hat{\sigma}_y\otimes\hat{\sigma}_y\right).
\end{align}
Here $0\leq\mathcal{C}_{A(B)}\leq 1$ with $\mathcal{C}_{A(B)}=1$ refers to a maximally entangled bipartite state (for example, a Bell state \cite{reid2009colloquium}) and $\mathcal{C}_{A(B)}=0$ indicates a fully separable (unentangled) state. 

For the present problem we introduce the basis set $\lbrace \ket{g_1g_2}\rightarrow\ket{1}, \ket{e_1g_2}\rightarrow\ket{2}, \ket{g_1e_2}\rightarrow\ket{3}, \ket{e_1e_2}\rightarrow\ket{4} \rbrace$. Next, subject to the initial condition $\hat{\rho}_{sys}(0)=\ket{g_1g_2}\bra{g_1g_2}$, we numerically solved the three-photon Fock state master equation. Therein, we find that the spin-flip density matrix of the two-QE system takes the following form with 8 out of 16 time-dependent density matrix elements remaining zero for all times
\begin{align}
\widetilde{\rho}_{12}(t)=
    \begin{pmatrix}
     \rho_1\rho_{16}+\rho^2_4 & 0 & 0 & \rho_1\rho_4\\
     0 & 2\rho^2_6 & 2\rho^2_6 & 0\\
     0 & 2\rho^2_6 & 2\rho^2_6 & 0\\
     \rho_1\rho_4 & 0 & 0 & \rho_{1}\rho_{16}
    \end{pmatrix}.
\end{align}
Note that we have adopted short notation here in which $\rho_1\equiv \bra{1}\hat{\rho}_{3,3}(t)\ket{1}$, $\rho_4\equiv \bra{1}\hat{\rho}_{3,3}(t)\ket{4}$, $\rho_6\equiv \bra{2}\hat{\rho}_{3,3}(t)\ket{2}$, and $\rho_{16}\equiv \bra{4}\hat{\rho}_{3,3}(t)\ket{4}$. Diagonalization of $\widetilde{\rho}_{12}(t)$ yields the following set of eigenvalues
\begin{align}
    &\lambda_1 = 0, ~~\lambda_2 = 4\rho^2_6, \nonumber\\
    &\lambda_3 = \rho_1\rho_{16}+\frac{1}{2}\rho_4\left(\rho_4-\sqrt{\rho^2_4+4\rho_1\rho_{16}}\right),~\text{and}\nonumber\\
    &\lambda_4 = \rho_1\rho_{16}+\frac{1}{2}\rho_4\left(\rho_4+\sqrt{\rho^2_4+4\rho_1\rho_{16}}\right).
\end{align}
Inserting these eigenvalues in Eq.~\eqref{concurdef}, one can find the entanglement between QEs. In Fig.~\ref{Fig3}(c) we plot this bipartite entanglement in both the bidirectional symmetric and chiral cases. In parts (a) and (b) of Fig.~\ref{Fig3} the populations corresponding to these two cases have also been plotted to aid the understanding of concurrence behavior. For the bidirectional symmetric case, we notice that the temporal profile of concurrence follows a pattern with two peaks (at $t=4.70\Gamma^{-1}$ and $t=6.65\Gamma^{-1}$) separated by a dip (centered at $t\sim 6\Gamma^{-1}$) while reaching the maximum value of up to $11\%$. The first peak is reached just before the three-photon wavepacket reaches its maximum value which can be argued to correspond to the partial formation of a Bell state involving both QEs excited i.e. $\left(\ket{g_1g_2}+\ket{e_1e_2}\right)/\sqrt{2}$ (as evident from the solid thin magenta curve of $P_{e_1e_2}$ in Fig.~\ref{Fig3}(a)). After that, as the wavepacket begins to leave the emitter region, one of the emitters decays hence now forming a Bell state like $\left(\ket{e_1g_2}+\ket{g_1e_2}\right)/\sqrt{2}$ which results in the increase in the concurrence again around $t= 6.65\Gamma^{-1}$ hence forming the second peak. Again the population plot (Fig.~\ref{Fig3}(a)) supports this explanation as $P_{e_1g_2+g_1e_2}$ (blue solid thick curve) decays slowly and dominates over $P_{e_1e_2}$ curve around $t\sim 7\Gamma^{-1}$ region.

In the chiral case ($\Gamma_{ir}=5\Gamma_{il}$, i=1,2), we find a marked change in the behavior of population and entanglement dynamics compared to the bidirectional symmetric case. On one hand, in Fig.~\ref{Fig3}(b) we observe $P_{e_1g_2+g_1e_2}$ (red solid curve) exhibiting a two-peak pattern with a maximum value increase by a factor of almost 2 compared to the symmetric coupling case (blue solid thick curve in Fig.~\ref{Fig3}(a)). On the other hand, the maximum value of both QEs excited probability $P_{e_1e_2}$ (brown solid thin curve in Fig.~\ref{Fig3}(b)) reduced more than 1/2 compared to the symmetric problem. We find that this single QE excited probability trend extended down to the entanglement behavior as well, where the concurrence in the chiral case (dashed red curve in Fig.~\ref{Fig3}(c)) showing a single peak pattern but with 5 times higher value achieved for the maximum entanglement. Furthermore, we note that chirality also assisted in sustaining this entanglement for times between $8\Gamma^{-1}$ to $10\Gamma^{-1}$ even after the three-photon wavepacket diminishes.

\subsection{\label{sec:level3.B}Three-QE case and tripartite entanglement}
\begin{figure}
\begin{center}
\begin{tikzpicture}
\draw[red, thick, dashed] (1,2) -- (7,2);
\draw[red, thick, dashed] (7,2) -- (4,-2);
\draw[red, thick, dashed] (4,-2) -- (1,2);
        
\filldraw[draw=black, fill=green!20, line width=1pt, dashed] (4,2) -- (4.75,-1) -- (1.75,1) -- cycle;
        
\node at (4.9, 0.5) {$C^2_{3(12)}$};
\node at (3.5, -0.65) {$C^2_{1(23)}$};
\node at (2.4, 1.7) {$C^2_{2(13)}$};
        
\filldraw[blue] (4,2) circle (3pt) node[anchor=south east]{QE-1};
\filldraw[blue] (4.75,-1) circle (3pt) node[anchor=north west]{QE-2};
\filldraw[blue] (1.75,1) circle (3pt) node[anchor=north east]{QE-3};
\end{tikzpicture}
\end{center}
\captionsetup{
  format=plain,
  margin=1em,
  justification=raggedright,
  singlelinecheck=false
}
\caption{(Color online) The concurrence triangle for a tripartite system (composed of QE-1, 2, and 3). Note that the length of each side of the triangle is equal to the square of the concurrence between different possible bipartite pairings.}\label{Fig4}
\end{figure}
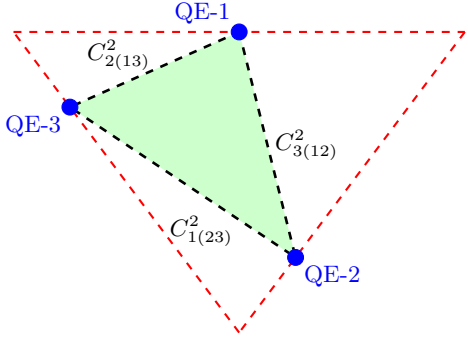
\begin{figure*}
\centering
\begin{tabular}{@{}cccc@{}}
\includegraphics[width=2.25in, height=1.75in]{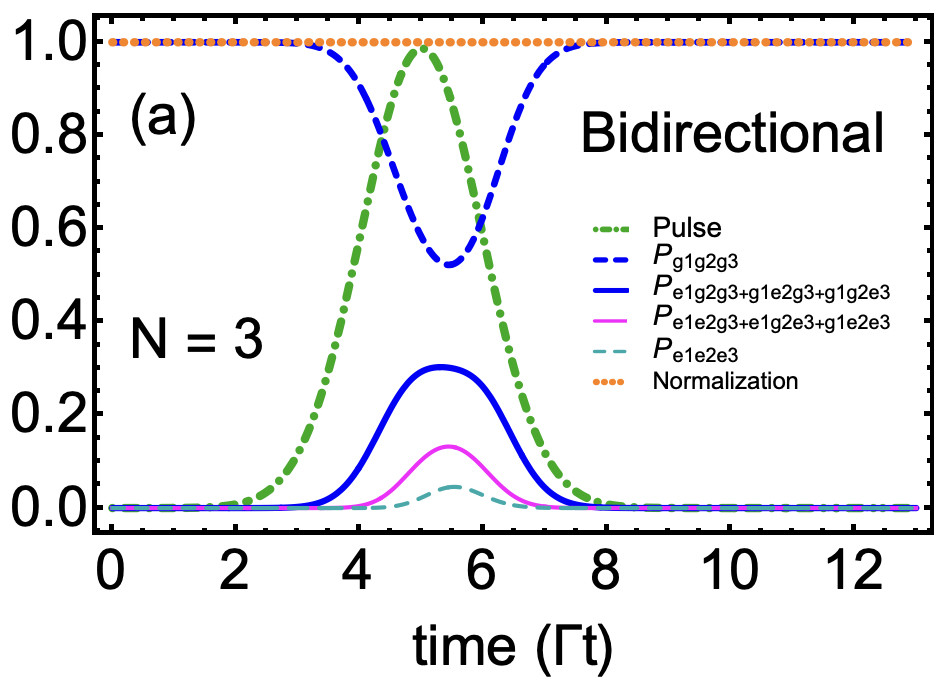} &
\hspace{-1.5mm}\includegraphics[width=2.25in, height=1.75in]{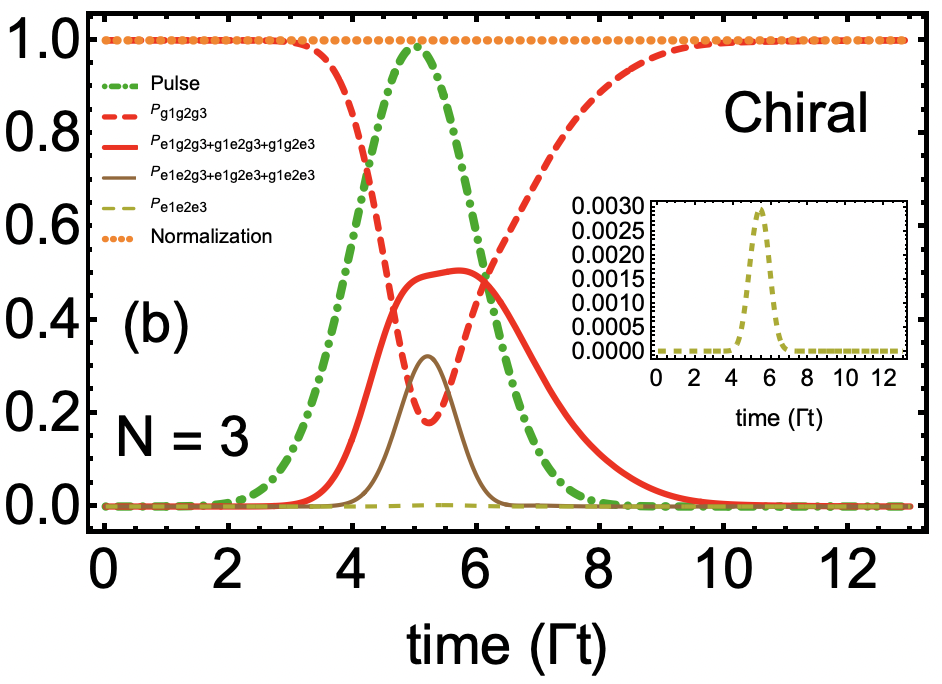}  &
\hspace{-1.5mm}\includegraphics[width=2.5in, height=1.75in]{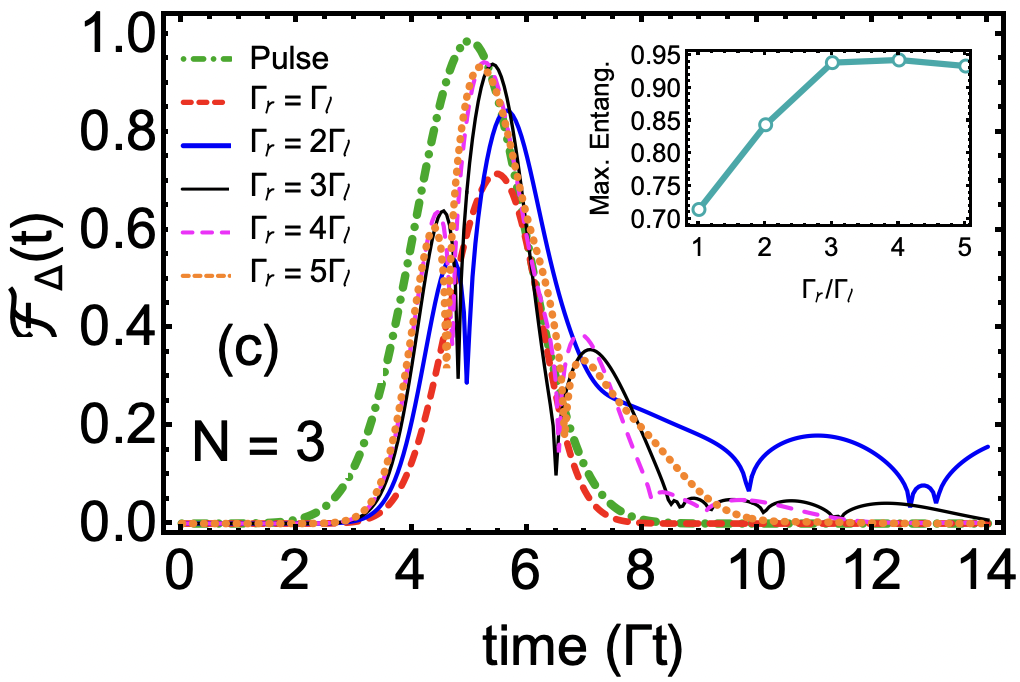}  
\end{tabular}
\captionsetup{
format=plain,
margin=1em,
justification=raggedright,
singlelinecheck=false
}
\caption{(Color online) Population dynamics for the three-photon three-QE ($N=3$) problem. (a) Symmetric bidirectional case i.e. $\Gamma_{ir}=\Gamma_{il}=1$; and (b) Chiral case with $\Gamma_{ir}/\Gamma_{il}=5$, $\forall i=1,2,3$. Similar to Fig.~\ref{Fig3}, all QEs are assumed to be identical and time delays have been ignored. In the plot legends, we are following the notation in which the first, second, and third slots in the subscripts represent the state of the first, second, and third QE, respectively. The inset in the plot (b) represents the curve of all three QEs being excited simultaneously ($P_{e_1e_2e_3}$). (c) Time evolution of tripartite entanglement among three QEs quantified through the concurrence fill $\mathcal{F}_\Delta(t)$ measure. Emitter-waveguide coupling strength in the right direction $\Gamma_r$ has been varied in units of $\Gamma_l$. The inset shows the behavior of maximum entanglement ($\mathcal{F}_{\Delta, max}$) achieved for each chosen value of $\Gamma_r/\Gamma_l$. The rest of the parameters in all plots are the same as used in Fig.~\ref{Fig2}.}\label{Fig5}
\end{figure*}

\begin{table*}
\centering
\caption{Maximum excitation probability comparison}
\begin{tabular}{c||c|c|c||c|c|cc}
\hline\hline
\multicolumn{1}{c||}{\textbf{Excitation}} & \multicolumn{3}{c||}{\textbf{~~Bidirectional~~}} & \multicolumn{4}{c}{\textbf{~~Chiral~~}} \\
\hline
& {N=1} & {N=2} & {N=3} & {N=1} & {N=2} & {N=3} &  \\
\hline
$P_{1,max}$ & $0.52$ \text{at} $5.25\Gamma^{-1}$ & $0.35$ \text{at} $5.47\Gamma^{-1}$ & $0.30$ \text{at} $5.31\Gamma^{-1}$ & $0.37$ \text{at} $4.85\Gamma^{-1}$ & $0.61$ \text{at} $6.57\Gamma^{-1}$ & $0.51$ \text{at} $5.71\Gamma^{-1}$\\
$P_{2,max}$ &  & $0.23$ \text{at} $5.44\Gamma^{-1}$ & 0.13 \text{at} $5.44\Gamma^{-1}$ &  & $0.08$ \text{at} $5.22\Gamma^{-1}$ & $0.32$ \text{at} $5.20\Gamma^{-1}$\\
$P_{3,max}$ &  &  & $0.05$ \text{at} $5.50\Gamma^{-1}$ &  & & $0.003$ \text{at} $5.41\Gamma^{-1}$\\
\hline\hline
\end{tabular}
\end{table*}

Moving on to the three-QE mixed states, it turns out that the bipartite concurrence measure doesn't extend down straightforwardly to the tripartite case \cite{yonacc2007pairwise, amico2008entanglement}. To this end, we apply a recently proposed tripartite entanglement measure by Xie and Eberly \cite{xie2021triangle}. This measure is reported to quantify genuine three-party entanglement by analyzing the area of the concurrence triangle (hence the name triangle measure or concurrence fill). The measure itself involves calculating the pairwise concurrence among all three QEs with a bipartite-split between $i$th qubit (treated as one subsystem) and $j$, $k$ qubit pair (as the other subsystem) as shown in Fig.~\ref{Fig4}). For the set of qubits $i,j,k$; such a ``one-to-other" concurrence is known to follow the identity \cite{zhu2015generalized}
\begin{align}
   \mathcal{C}^2_{i(jk)} \leq \mathcal{C}^2_{j(ki)} + \mathcal{C}^2_{k(ij)},
\end{align}
where, for example, $\mathcal{C}^2_{1(23)}$ is calculated using  \cite{woldekristos2009tripartite}. 
\begin{align}
    \mathcal{C}^2_{1(23)}=\sqrt{2(1-\tr\lbrace \hat{\rho}^2_1\rbrace)}~~ \text{with}~~ \hat{\rho}_1:=\tr_{23}\lbrace \hat{\rho}_{123} \rbrace.
    \end{align}
Here $\hat{\rho}_1$ represents the reduced density matrix of the first qubit obtained by tracing out the second and third qubit from the full system density matrix $\hat{\rho}_{123}$.
Thus, considering $\mathcal{C}^2_{1(23)}$, $\mathcal{C}^2_{2(31)}$, and $\mathcal{C}^2_{3(12)}$ as lengths of the side of a triangle, Xie and Eberly used Heron's expression for the area of such a triangle and arrived at the following formula that describes the triangle measure:
\begin{align}\label{FQmeas}
&\mathcal{F}_\Delta = \left[\frac{16}{3}
\mathcal{Q}\left(\mathcal{Q}-\mathcal{C}^2_{1(23)}\right)\left(\mathcal{Q}-\mathcal{C}^2_{2(13)}\right)
\left(\mathcal{Q}-\mathcal{C}^2_{3(12)}\right)\right]^{\frac{1}{4}},\nonumber\\
& \text{where}~~~\mathcal{Q}=\frac{1}{2}\left(\mathcal{C}^2_{1(23)}+\mathcal{C}^2_{2(13)}+\mathcal{C}^2_{3(12)}\right),
\end{align}
where the prefactor $\left(16/3\right)^{1/4}$ ensures that $\mathcal{F}_\Delta$ remains bounded between $0$ and $1$, again $1$ referring to the maximum of genuinely entangled tripartite state (such as W or GHZ state \cite{m2019tripartite}) and $0$ indicates a fully unentangled state. Furthermore, consistent with Fig.~\ref{Fig4}, $\mathcal{Q}$ is also called the half-perimeter of the concurrence triangle.\\

In Fig.~\ref{Fig5} we plot population and entanglement dynamics for the three-QEs problem. In Fig.~\ref{Fig5}(a) and Fig.~\ref{Fig5}(b) we compare the populations in bidirectional symmetric and chiral scenarios, respectively. With the presence of the third QE, all probabilities including single emitter being excited ($P_{e_1g_2g_3+g_1e_2g_3+g_1g_2e_3}$), double emitter excited ($P_{e_1e_2g_3+e_1g_2e_3+g_1e_2e_3}$) and triple emitter excited ($P_{e_1e_2e_3}$) have been reported. In both bidirectional and chiral scenarios, we note that as the number of excited QEs is increased the corresponding probability shows a considerable reduction. In particular, in the chiral case, $P_{e_1e_2e_3}$ becomes too tiny such that we have to include it as the inset in Fig.~\ref{Fig5}(b) where it reaches a maximum value of merely $0.3\%$. As summarized in Table 1, we find that the maximum value probability of one- ($P_{1,max}$), and two- ($P_{2,max}$) QE excited in the bidirectional model shows a noticeable decrease for $N=3$ case as compared to the respective $N=1$ and $N=2$ cases. However, in the chiral case, such a trend is broken. Additionally, by the comparison of Fig.~\ref{Fig3}(b) and Fig.~\ref{Fig4}(b), we notice that unlike $N=2$ problem with chiral couplings, $N=3$ chiral scenario fails to show any oscillatory behavior in the populations. But single excitation probability $P_{e_1g_2g_3+g_1e_2g_3+g_1g_2e_3}$ forms an almost plateau between $5\Gamma^{-1}\lesssim t \lesssim 6.5\Gamma^{-1} $ which helps $P_{e_1g_2g_3+g_1e_2g_3+g_1g_2e_3}$ to maintain a non-zero value for an additional $t\cong 1.5\Gamma^{-1}$ after the complete diminishing of the three-photon pulse. 

\begin{figure*}
\centering
\begin{tabular}{@{}cccc@{}}
\includegraphics[width=3.4in, height=2.25in]{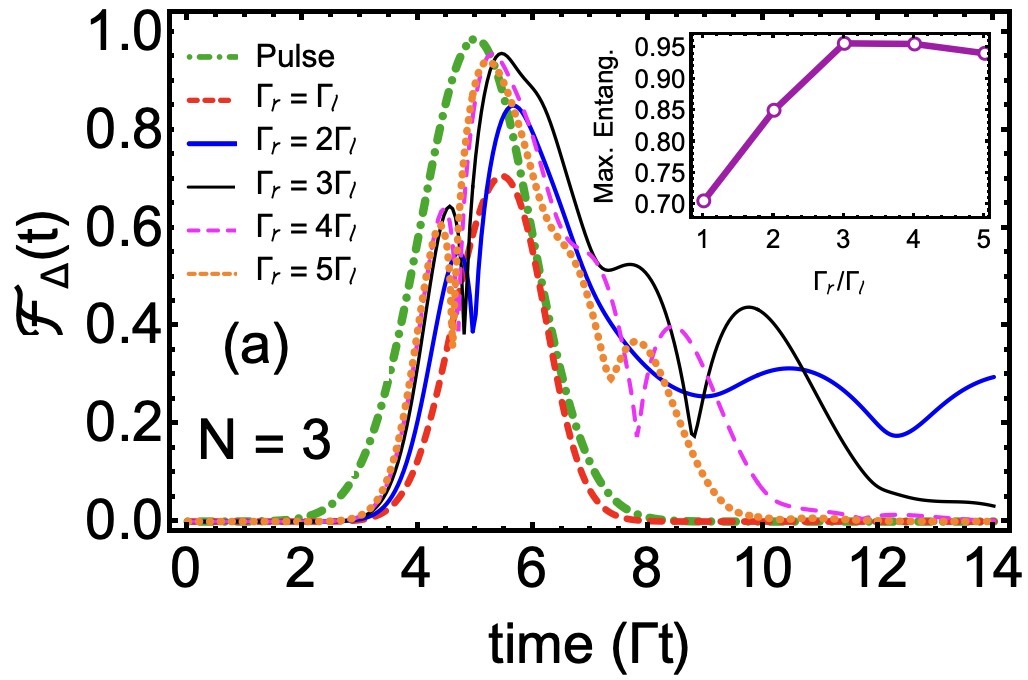} &
\includegraphics[width=3.4in, height=2.25in]{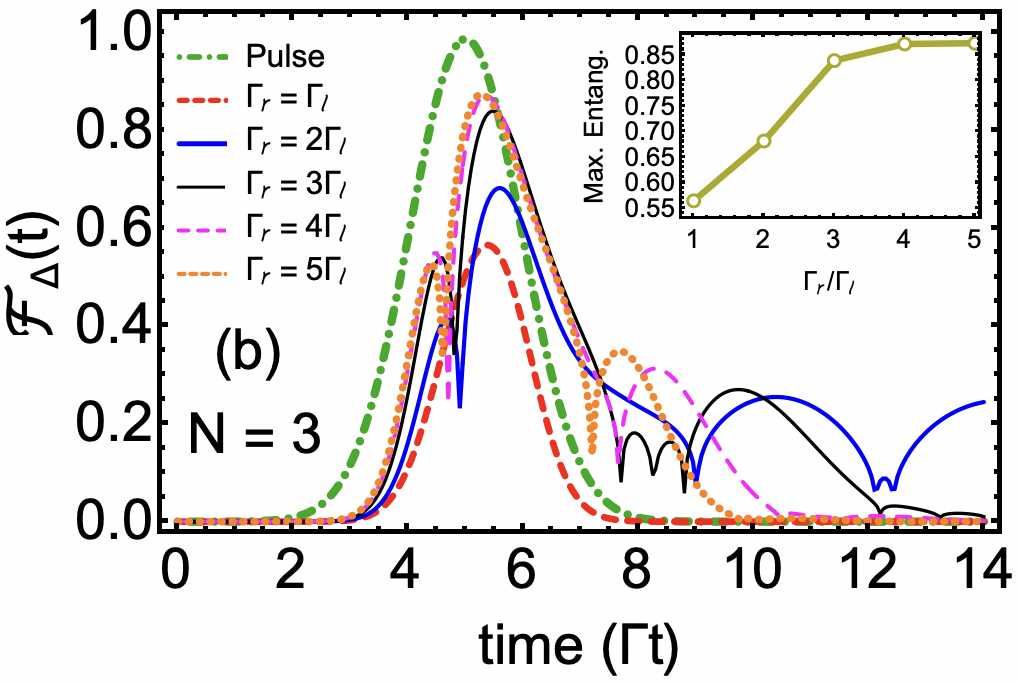}  
\end{tabular}
\captionsetup{
format=plain,
margin=1em,
justification=raggedright,
singlelinecheck=false
}
\caption{(Color online) (a) Time evolution of tripartite entanglement when all QEs' transition frequency is detuned by $\Gamma/2$ from the peak frequency of the three-photon wavepacket. Here we have set $\Gamma_l\equiv \Gamma$. (b) Entanglement dynamics in the presence of spontaneous emission rate $\gamma$ which is assumed to be the same for all QEs with a value of $3\Gamma/4$. Insets in both plots show the maximum entanglement as a function of $\Gamma_r$. Besides detuning and spontaneous emission rate, all parameters are the same as used previously. }\label{Fig7}
\end{figure*}

In Fig.~\ref{Fig5}(c) we plot the time evolution of concurrence fill while varying the right direction emitter-waveguide coupling $\Gamma_r$ (assumed to be the same for all QEs) from symmetric bidirectional case $\Gamma_r=\Gamma_l$ to the maximum chiral case $\Gamma_r=5\Gamma_l$. We notice, following the population trend observed in Fig.~\ref{Fig5}(a) and Fig.~\ref{Fig5}(b), for all non-chiral cases the entanglement among QEs survives for a time longer than the pulse duration. Additionally, the irregular oscillations in $\mathcal{F}_\Delta(t)$ for chiral case exhibit the phenomenon of entanglement collapse and revival \cite{mazzola2009sudden, xu2010experimental, xie2023evidence} which is more visible for the $\Gamma_r=3\Gamma_l$ case (thin blue curve). Most importantly, we notice that the maximum value achieved by the entanglement in all chiral cases poses an upper bound on the maximum value of entanglement achieved in the symmetric directional case where $\mathcal{F}_\Delta \cong 0.70$. This important finding is further emphasized in the inset plot in Fig.~\ref{Fig5}(c) where we observe this maximum value to be elevated by more than 35\% as we go from the symmetric bidirectional case of $\Gamma_r$ to chiral cases of $3\Gamma_l\leq\Gamma_r\leq 5\Gamma_l$. Note that for single-photon two-qubit wQED problem, Ballestero et al. have shown that the chirality can be used to enhance the maximum entanglement by a factor of 3/2 as compared to the corresponding symmetric bidirectional case \cite{gonzalez2015chiral}. Similarly, Mirza et al. (the corresponding author of this work) have reported the twice enhancement in qubit-qubit entanglement for the two-photon two-qubit case \cite{mirza2016two}. We on the other, in this work have shown that this trend extends down to genuine tripartite entanglement where $\Gamma_r\geq 3\Gamma_l$ case chirality assists to increase the concurrence fill among three-QEs by 35\% (factor of $\sim 5/14$).


\subsection{Tripartite entanglement in the presence of detuning and spontaneous emission}
\vspace{-3mm}
So far we have assumed an on-resonance scenario where the peak frequency of the three-photon wavepacket $\omega_p$ has been set equal to the emitter transition frequency $\omega_{eg}$. Additionally, we have completely ignored the photon emissions into non-waveguide modes through the process of spontaneous emission. We now address these two scenarios separately and plot the three-QE entanglement dynamics for a detuned case with no spontaneous emission (i.e. $\omega_p-\omega_{eg}=\Gamma/2$ and $\gamma=0$) in Fig.~\ref{Fig7}(a) and for an on-resonance case with a non-zero spontaneous emission scenario ($\omega_p=\omega_{eg}$ and $\gamma= 3\Gamma/4$) in Fig.~\ref{Fig7}(b).

From Fig.~\ref{Fig7}(a) we note that for all cases as we increase $\Gamma_r$ value from $\Gamma_l$ to $5\Gamma_l$, near the peak frequency of the wavepacket, detuning preserves the overall profile of the entanglement observed in the on-resonance situation. Additionally, from the inset plot, we notice that the maximum entanglement values also follow a quite similar pattern as found in the no-detuning problem. However, we observe the novel aspect of Fig.~\ref{Fig7}(a) in a long time ($t\gtrsim 8\Gamma^{-1}$) behavior of $\mathcal{F}_\Delta(t)$ where tripartite entanglement sustains for longer times and tend to produce more oscillatory behavior as compared to the no-detuning problem (compare, for instance, thin black ($\Gamma_r=3\Gamma_l$) curves in  Fig.\ref{Fig7}(a) and Fig.~\ref{Fig5}(c)). 

In Fig.~\ref{Fig7}(b) we study the impact of spontaneous emission on the tripartite entanglement under the strong coupling regime of wQED ($\gamma < \Gamma$). As expected, we find that the presence of a finite spontaneous emission considerably reduced the entanglement while keeping the overall profile of entanglement more or less the same. In particular, we point out that for $\gamma=3\Gamma/4$, the maximum value of entanglement for the symmetric bidirectional case shows a $15\%$ reduction compared to the $\gamma=0$ situation. Here we emphasize that the chirality not only assists to achieve elevated values of maximum entanglement in the presence of spontaneous emission but also helps to somewhat decrease the difference in the $\mathcal{F}_{\Delta, max}$ value (see for example, the most chiral situation of $\Gamma_r=5\Gamma_l$ in which the maximum entanglement difference reduces to $10\%$ compared to the corresponding $\gamma=0$ problem).


\section{Summary and Conclusions}\label{sec:IV}
In this paper, we studied the generation and control of three-photon Gaussian wavepacket-induced entanglement between 2 to 3 QEs side-coupled to chiral and symmetric bidirectional waveguides. Through the numerical solution of three-photon Fock state master equations, we calculated population dynamics and entanglement evolution which were quantified via bipartite concurrence and concurrence fill for two- and three-QE, cases respectively.

At the single QE level, we found that chiral light-matter interaction was able to achieve $\sim 37\%$ maximum excitation percentage probability which is smaller than $\sim 52\%$ percentage probability obtained for the bidirectional symmetric coupling case. However, starting from 2 QE case chirality began to exhibit considerable improvement in both gaining higher entanglement values as well as single-QE excitation probability. Particularly, for the set of parameters chosen in Fig.~3, we concluded that with a five times higher decay rate in the right waveguide direction compared to the left direction, a single QE excitation probability reaches values twice higher than the corresponding bidirectional symmetric cases. More importantly, this trend extends down to the emitter-emitter entanglement where the bipartite concurrence reached maximum values that were five times larger than the symmetric case.

For the $N=3$ QE problem, we found that for the bidirectional case, the maximum probability of single, double, and triply excited states show a considerable reduction as compared to the corresponding $N=1$ and $N=2$ problems. However, the chirality breaks this trend and also helps to sustain (at least) the single excitation probability (and hence the entanglement) for longer times. Furthermore, in the chiral case, we notice the phenomenon of tripartite entanglement death and revival. Importantly we point out that the maximum value achieved by the entanglement in all chiral cases (starting from $\Gamma_r=2\Gamma_l$ to $\Gamma_r=5\Gamma_l$) posed an upper bound on the maximum value of entanglement attained in the symmetric bidirectional problem ($\Gamma_r=\Gamma_l$). Compared to earlier studies of one and two-photon wQED where for two-qubit problem chirality is known to increase entanglement by a factor of 3/2 and 2, respectively; here for the three-photon case we have shown this enhancement to be $35\%$ (or by a factor of $\sim 5/14$).

Finally, we discuss the impact of detuning and spontaneous emission on the generated tripartite entanglement. There we concluded both small detunings ($\omega_p-\omega_{eg}=\Gamma/2$) and spontaneous emission rate ($\gamma=3\Gamma/4$) retain the overall temporal profile of the entanglement.  Detuning helps to sustain entanglement for longer times, while spontaneous emission rate results in a considerable reduction in the maximum value of entanglement. However, chirality still helped entanglement to show somewhat robustness against spontaneous emission loss. These behaviors convincingly show that three-photon chiral light-matter interactions can assist to accomplish higher maximum values of entanglement among QEs with better control to sustain genuine tripartite entanglement for elongated times.

\section*{Acknowledgements}
IMM would like to acknowledge financial support from the NSF Grant \# LEAPS-MPS 2212860 and the Miami University College of Arts and Science \& Physics Department start-up funding.

\setcounter{equation}{0}
\renewcommand\theequation{A.\arabic{equation}}
\section*{APPENDIX A. Quantum Langevin Equation for Cascaded Systems}
\subsection*{A.1. Quantum Langevin Equation for a single quantum system}
We begin by rewriting the total Hamiltonian $\hat{\mathcal{H}}$ (sum of Eq.~\eqref{Hsys}, Eq.~\eqref{Hbath}, and Eq.~\eqref{Hint}) without specifying $\hat{H}_{sys}$ for short notation and using $\hat{c}=\hat{\sigma}_1$ while focuing on a single QE in the emitter chain
\begin{align}
&\hat{\mathcal{H}} = \hat{\mathcal{H}}_{sys}+\int\limits^\infty_{-\infty}\hbar\omega\hat{b}^\dagger_r(\omega)\hat{b}_r(\omega)d\omega+\int\limits^\infty_{-\infty}\hbar\nu\hat{b}^\dagger_l(\nu)\hat{b}_l(\nu)d\nu\nonumber\\
&-i\hbar\sqrt{\frac{\Gamma_{1r}}{2\pi}}\int\limits^\infty_{-\infty}\Big(e^{ik_0d_1}\hat{c}^\dagger_1\hat{b}_r(\omega)-e^{-ik_0d_1}\hat{b}^\dagger_r(\omega)\hat{c}_1\Big)d\omega\nonumber\\
&-i\hbar\sqrt{\frac{\Gamma_{1l}}{2\pi}}\int\limits^\infty_{-\infty}\Big(e^{-ik_0d_1}\hat{c}^\dagger_1\hat{b}_l(\nu)-e^{ik_0d_1}\hat{b}^\dagger_l(\nu)\hat{c}_1\Big)d\nu.
\end{align}
In the Heisenberg picture, the equations of motion for the right continuum, for the left continuum, and for an arbitrary operator $\hat{\mathcal{X}}_1(t)$ (which may or may not be $\hat{c}_1(t)$) are given by
\begin{subequations}
\begin{align}
&\frac{d\hat{b}_r(\omega; t)}{dt}=-i\omega\hat{b}_r(\omega;t)+\sqrt{\frac{\Gamma_{1r}}{2\pi}}e^{-ik_0d_1}\hat{c}_1,\label{dbR}\\
&\frac{d\hat{b}_l(\nu; t)}{dt}=-i\omega\hat{b}_l(\nu;t)+\sqrt{\frac{\Gamma_{1l}}{2\pi}}e^{ik_0d_1}\hat{c}_1,\label{dbL}\\
&\frac{d\hat{\mathcal{X}}_1(t)}{dt}=-\frac{i}{\hbar}\left[\hat{\mathcal{X}}_1,\hat{\mathcal{H}}_{sys}\right]\nonumber\\
&-\sqrt{\frac{\Gamma_{1r}}{2\pi}}\int\limits^\infty_{-\infty}\Big(e^{ik_0d_1}\left[\hat{\mathcal{X}}_1,\hat{c}^\dagger_1\right]\hat{b}_r(\omega;t)- h.c.\Big)d\omega\nonumber\\
&-\sqrt{\frac{\Gamma_{1l}}{2\pi}}\int\limits^\infty_{-\infty}\Big(e^{-ik_0d_1}\left[\hat{\mathcal{X}}_1,\hat{c}^\dagger_1\right]\hat{b}_l(\nu;t)-h.c.\Big)d\nu.\label{dX1}
\end{align}
\end{subequations}
To eliminate continua from Eq.~\eqref{dX1} we integrate Eq.~\eqref{dbR} and Eq.~\eqref{dbL} from some initial time $t_0$ to present time $t$ to find
\begin{subequations}\label{bRbLSols}
\begin{align}
\hat{b}_r(\omega;t)&= \hat{b}_r(\omega;t_0)e^{-i\omega(t-t_0)}\nonumber\\
&+\sqrt{\frac{\Gamma_{1r}}{2\pi}}e^{-ik_0d_1}\int\limits^t_{t_0}\hat{c}_1(t^{'})e^{-i\omega(t-t^{'})}dt^{'},\\
\hat{b}_l(\nu;t)&=\hat{b}_l(\nu;t_0)e^{-i\nu(t-t_0)}\nonumber\\
&+\sqrt{\frac{\Gamma_{1l}}{2\pi}}e^{ik_0d_1}\int\limits^t_{t_0}\hat{c}_1(t^{'})e^{-i\nu(t-t^{'})}dt^{'}.
\end{align}
\end{subequations}
Inserting Eq.~\eqref{bRbLSols} into Eq.~\eqref{dX1} and performing the integrations we arrive at 
\begin{widetext}
\begin{align}\label{QLanEq}
&\frac{d\hat{\mathcal{X}}_1(t)}{dt}=\frac{-i}{\hbar}\left[\hat{\mathcal{X}}_1(t),\hat{\mathcal{H}}_{sys}\right]-\left[\hat{\mathcal{X}}_1(t),\hat{c}^\dagger_1(t)\right]\left\lbrace\sqrt{\Gamma_{1r}}~e^{ik_0d_1}\hat{b}^{(1r)}_{in}(t)+\sqrt{\Gamma_{1l}}~e^{-ik_0d_1}\hat{b}^{(1l)}_{in}(t)+\left(\frac{\Gamma_{1r}+\Gamma_{1l}}{2}\right)\hat{c}_1(t)\right\rbrace\nonumber\\
&+\left\lbrace\sqrt{\Gamma_{1r}}~e^{-ik_0d_1}\hat{b}^{\dagger(1r)}_{in}(t)+\sqrt{\Gamma_{1l}}~e^{ik_0d_1}\hat{b}^{\dagger(1l)}_{in}(t)+\left(\frac{\Gamma_{1r}+\Gamma_{1l}}{2}\right)\hat{c}^\dagger_1(t)\right\rbrace\left[\hat{\mathcal{X}}_1(t),\hat{c}_1(t)\right].
\end{align}
\end{widetext}
This is the quantum Langevin equation \cite{gardiner2004quantum, gheri1998photon} describing the open system dynamics in the Heisenberg picture. Note that in deriving this equation we have defined two operators
\begin{subequations}
\begin{align}
&\int\limits^{+\infty}_{-\infty}\hat{b}_r(\omega;t_0)e^{-i\omega(t-t_0)}d\omega:=\sqrt{2\pi}~\hat{b}^{(1r)}_{in}(t),~~\text{and}\\
&\int\limits^{+\infty}_{-\infty}\hat{b}_l(\nu;t_0)e^{-i\nu(t-t_0)}d\nu:=\sqrt{2\pi}~\hat{b}^{(1l)}_{in}(t).   
\end{align}
\end{subequations}
These are the so-called input operators which, with a factor of $\sqrt{\Gamma_{1r}}$ and $\sqrt{\Gamma_{1l}}$ describe the impact of quantum noise on the system dynamics at initial times. We also notice that these input operators obey the following commutation relations to ensure causality:
\begin{align}
&\Big[\hat{b}^{(1r)}_{in}(t),\hat{b}^{\dagger(1r)}_{in}(t^{'})\Big]=\delta(t-t^{'}),~~\text{and}\nonumber\\
&\Big[\hat{b}^{(1l)}_{in}(t),\hat{b}^{\dagger(1l)}_{in}(t^{'})\Big]=\delta(t-t^{'}).
\end{align}
The remaining terms in Eq.~\eqref{QLanEq} have the following interpretation. The first commutator term describes the closed system dynamics (obtained from the standard Heisenberg equation of motion) while the terms with prefactors $\Gamma_{1r}$ and $\Gamma_{1l}$ represent an irreversible loss of energy from the system into the environmental degrees of freedom in the right and left direction, respectively.  

\subsection*{A.2. Input-output relations for a single quantum system}
Next, we take the equation of motion for the continua operators, and rather than integrating from past time $t_0$ to present time $t$ we now integrate the equations from present time $t$ to some future time $t_1$. Integrating over frequencies we obtain the following set of equations
\begin{subequations}
\begin{align}
&\int\limits^\infty_{-\infty}\hat{b}_r(\omega;t)d\omega=\sqrt{2\pi}~\hat{b}^{(1r)}_{out}(t)-\sqrt{\frac{\pi\Gamma_{1r}}{2}}e^{-ik_0d_1}\hat{c}_1(t),\\
& \int\limits^\infty_{-\infty}\hat{b}_l(\nu;t)d\nu=\sqrt{2\pi}~\hat{b}^{(1l)}_{out}(t)-\sqrt{\frac{\pi\Gamma_{1l}}{2}}e^{ik_0d_1}\hat{c}_1(t),
\end{align}
\end{subequations}
where we have now defined the output operators $\sqrt{2\pi}~\hat{b}^{(1r)}_{out}:=\int^\infty_{-\infty}\hat{b}_r(\omega;t_1)e^{-i\omega(t-t_1)}d\omega$ and $\sqrt{2\pi}~\hat{b}^{(1l)}_{out}:=\int^\infty_{-\infty}\hat{b}_l(\nu;t_1)e^{-i\nu(t-t_1)}d\nu$. When we compare the above set of equations with the past time $t_0$ version of the frequency-integrated equations of motion for continua we find
\begin{subequations}\label{InOut1}
\begin{align}
&\hat{b}^{(1r)}_{out}(t)=\hat{b}^{(1r)}_{in}(t)+\sqrt{\Gamma_{1r}}~e^{-ik_0d_1}\hat{c}_1(t),\\
&\hat{b}^{(1l)}_{out}(t)=\hat{b}^{(1l)}_{in}(t)+\sqrt{\Gamma_{1l}}~e^{ik_0d_1}\hat{c}_1(t).
\end{align}
\end{subequations}
Eq.~\eqref{InOut1} are the input-output relations of Collett and Gardiner \cite{gardiner1985input} now derived for the wQED where the phase factors indicate the location of the QE. These relations develop a connection between the input and output operators while incorporating the system's response to the input field.

\subsection*{A.3. Inclusion of a second quantum system: quantum Langevin equation for cascaded systems}
Extending our calculations to include the second QE into our model, we note that the output of the first (second) QE serves as the input to the second (first) QE in the right (left) direction. In this situation, our input-output relations take the form
\begin{subequations}
\begin{align}
\hat{b}^{(2r)}_{in}(t)&=\hat{b}^{(1r)}_{out}(t-\tau)=\hat{b}^{(1r)}_{in}(t-\tau)\nonumber\\
&+\sqrt{\Gamma_{1r}}e^{-ik_0d_1}\hat{c}_1(t-\tau),\\
\hat{b}^{(1l)}_{in}(t)&=\hat{b}^{(2l)}_{out}(t-\tau)=\hat{b}^{(2l)}_{in}(t-\tau)\nonumber\\
&+\sqrt{\Gamma_{2l}}e^{ik_0d_2}\hat{c}_2(t-\tau), \end{align}
\end{subequations}
where $\tau=L/c$ (time delay) represents the time taken by the photon traveling between the two QEs separated by a distance $L$ with the group velocity $c$ in the waveguide medium. Similar to Eq.~\eqref{QLanEq}, the quantum Langevin equation for the second QE, with some arbitrary operator $\hat{\mathcal{X}}_2(t)$ is given by
\begin{widetext}
\begin{align}\label{QLanEq2}
&\frac{d\hat{\mathcal{X}}_2(t)}{dt}=\frac{-i}{\hbar}\left[\hat{\mathcal{X}}_2(t),\hat{\mathcal{H}}_{sys}\right]-\left[\hat{\mathcal{X}}_2(t),\hat{c}^\dagger_2(t)\right]\left\lbrace\sqrt{\Gamma_{2r}}~e^{ik_0d_2}\hat{b}^{(2r)}_{in}(t)+\sqrt{\Gamma_{2l}}~e^{-ik_0d_2}\hat{b}^{(2l)}_{in}+\left(\frac{\Gamma_{2r}+\Gamma_{2l}}{2}\right)\hat{c}_2(t)\right\rbrace\nonumber\\
&+\left\lbrace\sqrt{\Gamma_{2r}}~e^{-ik_0d_2}\hat{b}^{\dagger(2r)}_{in}(t)+\sqrt{\Gamma_{2l}}~e^{ik_0d_2}\hat{b}^{\dagger(2l)}_{in}(t)+\left(\frac{\Gamma_{2r}+\Gamma_{2l}}{2}\right)\hat{c}^\dagger_2(t)\right\rbrace\left[\hat{\mathcal{X}}_2(t),\hat{c}_2(t)\right].
\end{align}
\end{widetext}
Next, to derive the quantum Langevin equation for cascaded systems \cite{carmichael2009statistical, daley2014quantum}, we suppose an arbitrary operator $\hat{\mathcal{X}}(t)$ which either belongs to system-1 or to the system-2. Thus, for $\hat{\mathcal{X}}(t)$ Eq.~\eqref{QLanEq} and Eq.~\eqref{QLanEq2} can be combined into a single equation as
\begin{widetext}
\begin{align}\label{QLanEq1+2}
&\frac{d\hat{\mathcal{X}}(t)}{dt}=\frac{-i}{\hbar}\left[\hat{\mathcal{X}}(t),\hat{\mathcal{H}}_{sys}\right]-\left[\hat{\mathcal{X}}(t),\hat{c}^\dagger_1(t)\right]\left\lbrace\sqrt{\Gamma_{1r}}~e^{ik_0d_1}\hat{b}^{(1r)}_{in}(t)+\sqrt{\Gamma_{1l}}~e^{-ik_0d_1}\hat{b}^{(1l)}_{in}(t)+\left(\frac{\Gamma_{1r}+\Gamma_{1l}}{2}\right)\hat{c}_1(t)\right\rbrace\nonumber\\
&+\left\lbrace\sqrt{\Gamma_{1r}}~e^{-ik_0d_1}\hat{b}^{\dagger(1r)}_{in}(t)+\sqrt{\Gamma_{1l}}~e^{ik_0d_1}\hat{b}^{\dagger(1l)}_{in}(t)+\Big(\frac{\Gamma_{1r}+\Gamma_{1l}}{2}\Big)\hat{c}^\dagger_1(t)\right\rbrace\left[\hat{\mathcal{X}}(t),\hat{c}_1(t)\right]-\left[\hat{\mathcal{X}}(t),\hat{c}^\dagger_2(t)\right]\bigg\lbrace\sqrt{\Gamma_{2r}}~e^{ik_0d_2}\nonumber\\
&\times \hat{b}^{(2r)}_{in}(t)+\sqrt{\Gamma_{2l}}~e^{-ik_0d_2}\hat{b}^{(2l)}_{in}(t)+\left(\frac{\Gamma_{2r}+\Gamma_{2l}}{2}\right)\hat{c}_2(t)\bigg\rbrace+\bigg\lbrace\sqrt{\Gamma_{2r}}~e^{-ik_0d_2}\hat{b}^{\dagger(2r)}_{in}(t)+\sqrt{\Gamma_{2l}}~e^{ik_0d_2}\hat{b}^{\dagger(2l)}_{in}(t)\nonumber\\
&+\Big(\frac{\Gamma_{2r}+\Gamma_{2L}}{2}\Big)\hat{c}^\dagger_2(t)\bigg\rbrace\left[\hat{\mathcal{X}}(t),\hat{c}_2(t)\right].
\end{align}
\end{widetext}
Moving forward, we connect the output of one QE to the input of the other. To this end, we set the time delay $\tau=0$ under the assumption $\omega_{eg_{j}},\Gamma_{jr},\Gamma_{jl}\ll 1/\tau=c/L$ (the system dynamics occurs on a time scale much longer than the time taken by the photons to propagate from one QE to another). We thus find the quantum Langevin equation for the two QEs cascaded through the bidirectional waveguide as
\begin{widetext}
\begin{align}\label{QLanEq2Fin}
&\frac{d\hat{\mathcal{X}}(t)}{dt}=\frac{-i}{\hbar}\left[\hat{\mathcal{X}}(t),\hat{\mathcal{H}}_{sys}\right]-\sum\limits^N_{j=1}\left[\hat{\mathcal{X}}(t),\hat{c}^\dagger_j(t)\right]\left\lbrace\sqrt{\Gamma_{jr}}~e^{ik_0d_j}\hat{b}^{(jr)}_{in}(t)+\sqrt{\Gamma_{jl}}~e^{-ik_0d_j}\hat{b}^{(jl)}_{in}(t)+\left(\frac{\Gamma_{jr}+\Gamma_{jl}}{2}\right)\hat{c}_j(t)\right\rbrace\nonumber\\
&+\sum^N_{j=1}\bigg\lbrace\sqrt{\Gamma_{jr}}~e^{-ik_0d_j}\hat{b}^{\dagger(jr)}_{in}(t)+\sqrt{\Gamma_{jl}}~e^{ik_0d_j}\hat{b}^{\dagger(jl)}_{in}(t)+\left(\frac{\Gamma_{jr}+\Gamma_{jl}}{2}\right)\hat{c}^\dagger_j(t)\bigg\rbrace \left[\hat{\mathcal{X}}(t),\hat{c}_j(t)\right]-\sqrt{\Gamma_{1l}\Gamma_{2l}}\Bigg(\left[\hat{\mathcal{X}}(t),\hat{c}^\dagger_1(t)\right]\nonumber\\
&\times e^{ik_0(d_2-d_1)}\hat{c}_2(t)-e^{-ik_0(d_2-d_1)}\hat{c}^\dagger_2(t)\left[\hat{\mathcal{X}}(t),\hat{c}_1(t)\right]\Bigg)-\sqrt{\Gamma_{1r}\Gamma_{2r}}\Bigg(\left[\hat{\mathcal{X}}(t),\hat{c}^\dagger_2(t)\right]e^{ik_0(d_2-d_1)}\hat{c}_1(t)-e^{-ik_0(d_2-d_1)}\hat{c}^\dagger_1(t)\nonumber\\
&\times\left[\hat{\mathcal{X}}(t),\hat{c}_2(t)\right]\Bigg).
\end{align}
\end{widetext}
\vspace{-10mm}

\subsection*{A.4. Extension to {\it N} quantum systems and three-photon Fock state master equation}
Through the inspection of the structure of Eq.~\eqref{QLanEq2Fin}, one can extend the problem to $N$ number of quantum systems, which results in the following quantum Langevin equation
\begin{widetext}
\begin{align}\label{QLanEq2Fin2}
&\frac{d\hat{\mathcal{X}}(t)}{dt}=\frac{-i}{\hbar}\left[\hat{\mathcal{X}}(t),\hat{\mathcal{H}}_{sys}\right]-\sum\limits^N_{j=1}\left\lbrace\left[\hat{\mathcal{X}}(t),\hat{c}^\dagger_j(t)\right]\Big\lbrace\sqrt{\Gamma_{jr}}~e^{ik_0d_j}\hat{b}^{(jr)}_{in}(t)+\sqrt{\Gamma_{jl}}~e^{-ik_0d_j}\hat{b}^{(jl)}_{in}(t)+\left(\frac{\Gamma_{jr}+\Gamma_{1l}}{2}\right)\hat{c}_j(t)\right\rbrace\nonumber\\
&-\Big\lbrace\sqrt{\Gamma_{jr}}~e^{-ik_0d_j}\hat{b}^{\dagger(jr)}_{in}(t)+\sqrt{\Gamma_{jl}}~e^{ik_0d_j}\hat{b}^{\dagger(jl)}_{in}(t)+\left(\frac{\Gamma_{jr}+\Gamma_{jl}}{2}\right)\hat{c}^\dagger_j(t)\Big\rbrace\left[\hat{\mathcal{X}}(t),\hat{c}_j(t)\right]\Bigg\rbrace-\sum_{\substack{j,m=1,\\j\neq m}}^{N}\Bigg\lbrace\left[\hat{\mathcal{X}}(t),\hat{c}^\dagger_j(t)\right]\nonumber\\
&\times e^{ik_0(d_j-d_m)}\Big(\sqrt{\Gamma_{jr}\Gamma_{mr}}\hat{c}_m(t)\delta_{j>m}+\sqrt{\Gamma_{jl}\Gamma_{ml}}\hat{c}_m(t)\delta_{j<m}\Big)-e^{-ik_0(d_j-d_m)}\Big(\sqrt{\Gamma_{jr}\Gamma_{mr}}\hat{c}^\dagger_m(t)\delta_{j>m}+\sqrt{\Gamma_{jl}\Gamma_{ml}}\nonumber\\
&\times \hat{c}^\dagger_m(t)\delta_{j<m}\Big)\left[\hat{\mathcal{X}}(t),\hat{c}_j(t)\right]\Bigg\rbrace.
\end{align}
\end{widetext}
Next, to transform into the Schr\"odinger picture we use the identity
\begin{align}
\tr_{S\oplus R}\left\lbrace \frac{d\hat{\mathcal{X}}(t)}{dt}\hat{\rho}(t_0)\right\rbrace=\tr_S\left\lbrace \hat{\mathcal{X}}(t_0)\frac{d\hat{\rho}_S(t)}{dt}\right\rbrace,
\end{align}
where $S$ and $R$ here stand for the system and reservoir/environment degrees of freedom, respectively. $\hat{\rho}_S(t)$ is the system density operator we are trying to find out. To this end, we use the above-mentioned identity and the cyclic property of trace we arrive at the following master equations applicable to the waveguide QED architecture under study
\begin{widetext}
\begin{align}\label{masEqF}
&\frac{d\hat{\rho}_S(t)}{dt}=\frac{-i}{\hbar}\left[\hat{\mathcal{H}}_{sys},\hat{\rho}_S(t)\right]-\sum\limits^N_{j=1}\left(\frac{\Gamma_{jr}+\Gamma_{jl}}{2}\right)\bigg(\hat{c}^\dagger_j\hat{c}_j\hat{\rho}_S(t)-2\hat{c}_j\hat{\rho}_S(t)\hat{c}^\dagger_j+\hat{\rho}_S(t)\hat{c}^\dagger_j\hat{c}_j \bigg)-\sum_{\substack{j,m=1,\\j\neq m}}^{N}\bigg(\sqrt{\Gamma_{jr}\Gamma_{mr}}\delta_{j>m}\nonumber\\
&+\sqrt{\Gamma_{jl}\Gamma_{ml}}\delta_{j<m} \bigg)\bigg\lbrace e^{ik_0(d_j-d_m)}\left(\hat{c}^\dagger_j\hat{c}_m\hat{\rho}_S(t)-\hat{c}_m\hat{\rho}_S(t)\hat{c}^\dagger_j\right)-e^{-ik_0(d_j-d_m)}\left(\hat{c}_j\hat{\rho}_S(t)\hat{c}^\dagger_m-\hat{\rho}_S(t)\hat{c}^\dagger_m\hat{c}_j\right)\bigg\rbrace\nonumber\\
&-\sum^N_{j=1}\sqrt{\Gamma_{jr}}e^{ik_0d_j}\Big[\hat{c}^\dagger_j\tr_R\Big\lbrace\hat{U}(t-t_0)\hat{b}^{(jr)}_{in}(t)\hat{\rho}(t_0)\hat{U}^\dagger(t-t_0)\Big\rbrace-\tr_R\Big\lbrace\hat{U}(t-t_0)\hat{b}^{(jr)}_{in}(t)\hat{\rho}(t_0)\hat{U}^\dagger(t-t_0)\Big\rbrace\hat{c}^\dagger_j\Big]\nonumber\\
&+\sum^N_{j=1}\sqrt{\Gamma_{jr}}e^{-ik_0d_j}\Big[\hat{c}_j\tr_R\Big\lbrace\hat{U}(t-t_0)\hat{\rho}(t_0)\hat{b}^{\dagger(jr)}_{in}(t)\hat{U}^\dagger(t-t_0)\Big\rbrace-\tr_R\Big\lbrace\hat{U}(t-t_0)\hat{\rho}(t_0)\hat{b}^{\dagger(jr)}_{in}(t)\hat{U}^\dagger(t-t_0)\Big\rbrace\hat{c}_j\Big].
\end{align}
\end{widetext}
Here we have made a considerable simplification by observing that the right end of the waveguide is driven by a vacuum state which leads to the vanishing of the left direction input term as
\begin{align}
&\tr_{S\oplus R}\left\lbrace\left[\hat{\mathcal{X}}(t),\hat{c}^\dagger_j(t)\right]\hat{b}^{(jl)}_{in}(t)\hat{\rho}(t_0) \right\rbrace = 0.
\end{align}
Note that, on the other hand, input terms for the left end of the waveguide do not vanish due to the presence of a three-photon wavepacket input. This, for example, for $j=1$ case gives us 
\begin{align}
\hat{b}^{(1r)}_{in}(t)\hat{\rho}(t_0)=&\sqrt{3}\sum_{\alpha,\beta,\mu}g_{\alpha,\beta,\mu}\bigotimes_j\ket{g_j}\otimes\widetilde{f}_\alpha(t)\nonumber\\
&\times\ket{\Psi_{\beta\mu}}\bra{\Psi_r}\otimes\ket{vac}_l\bra{vac},
\end{align}
where $\bigotimes_j\ket{g_j}\equiv \ket{g_1}\otimes\ket{g_2}...\ket{g_N}$ showing our choice of initial condition in which all QEs are in their ground state. $g_{\alpha\beta\mu}$ and $\widetilde{f}_\alpha(t)$ is the coefficient and Fourier transform of the function $f_\alpha(\omega)$, respectively both used in the expansion of the spectral density $\mathcal{G}(\omega_1,\omega_2,\omega_3)$. Explicitly $\widetilde{f}_\alpha(t)$ is given by
\begin{align*}
\widetilde{f}_\alpha(t)=\frac{1}{\sqrt{2\pi}}\int\limits^{+\infty}_{-\infty} e^{-i\omega(t-t_0)}f_\alpha(\omega)d\omega,~\text{while setting}~t_0=0.
\end{align*}
And $\ket{\Psi_{\beta\mu}}$ is the two-photon state generated due to the annihilation of a single photon from the initial three-photon wavepacket and it takes the form
\begin{align*}
\ket{\Psi_{\beta\mu}}=\frac{1}{\sqrt{2}}\int\limits^{+\infty}_{-\infty}\int\limits^{+\infty}_{-\infty} f_\beta(\omega_2)f_\mu(\omega_3)\hat{b}^\dagger_r(\omega_2)\hat{b}^\dagger_r(\omega_3)\ket{vac}d\omega_2 d\omega_3.
\end{align*}
Finally, using the master equation given in Eq.~\eqref{masEqF}, along with these initial conditions, the dynamics of our system can be calculated under such three-photon Fock state master equations \cite{patrick2023fock}. Note that in order to obtain a closed set of differential equations describing the evolution of the emitter chain one would need to define new operators (such as $\hat{\rho}_{\beta\mu 3}:=\ket{\Psi_{\beta\mu}}\bra{\Psi_r}$). The equation of motion for such operators can be obtained using the general identity
\begin{align}
\tr_{S\oplus R}\left\lbrace \frac{d\hat{\mathcal{X}}}{dt}\varpi(t_0)\right\rbrace=\tr_S\left\lbrace \hat{\mathcal{X}}(t_0)\frac{d\varpi(t)}{dt}\right\rbrace.
\end{align}

\bibliographystyle{ieeetr}
\bibliography{paper}
\end{document}